\documentclass[twocolappendix]{./emulateapj} %Looks like ApJ. 
\usepackage{amsmath,amssymb,amsfonts} 
\usepackage{color}                    % For creating coloured text and background
\usepackage{epsfig}
\usepackage{amsbsy} 
\usepackage{natbib}
\newcommand{\beq}	{\begin{equation}}
\newcommand{\eeq}	{\end{equation}}
\newcommand{\beqa}{\begin{eqnarray}}
\newcommand{\eeqa}{\end{eqnarray}}
\newcommand{\beqs}	{\begin{displaymath}}
\newcommand{\eeqs}	{\end{displaymath}}
\newcommand{\beqas}	{\begin{eqnarray*}}
\newcommand{\eeqas}	{\end{eqnarray*}}
 
\newcommand{\dis}{\displaystyle}
\def\bit{\begin{itemize}}
\def\eit{\end{itemize}}
\def\simlt{\lower.5ex\hbox{$\; \buildrel < \over \sim \;$}}
\def\simgt{\lower.5ex\hbox{$\; \buildrel > \over \sim \;$}}
\def\la{\simlt}
\def\ga{\simgt}

\def\Mc{M_{\rm c}}
\def\Mg{M_{\rm G}}
\def\Mb{M_{\rm B}}

\newcommand{\Mach}{{\cal M}}
\newcommand{\Alfven}{Alfv\'{e}n\ }

\def\Ramses{\texttt{RAMSES}\ }
\def\Orion{\texttt{ORION2}\ }

\def\b{\beta}

\def\calm  {{\cal M}}

\def\tb{t_{\rm B}}
\def\cs{c_{\rm s}}
\def\p{\partial}

\newcommand{\bvv}{\boldsymbol{\rm v}}
\newcommand{\bB}{\boldsymbol{\rm B}}
\newcommand{\bx}{\boldsymbol{\rm x}}

\newcommand{\ma}		{\calm_{\rm A}}

\newcommand{\rb}		{{r_{\rm B}}}

\newcommand{\rbh}		{{r_{\rm BH}}}
\newcommand{\rabh}		{{r_{\rm ABH}}}

\newcommand{\ravg}		{r_{\rm avg}}
\newcommand{\va}		{v_{\rm A}}
\newcommand{\vao}		{v_{\rm A,0}}

\newcommand{\vf}		{v_{\rm F}}

\newcommand{\mdotb} {\dot{M}_{\rm B}}
\newcommand{\avir}		{\alpha_{\rm vir}}

\newcommand{\e}	{$^{-1}$}

\newcommand{\eee}	{$^{-3}$}
\def\beq{\begin{equation}}
\def\eeq{\end{equation}}

\slugcomment{accepted to ApJ January 4th, 2014}

\shorttitle{Magnetic Bondi-Hoyle Accretion}
\shortauthors{Lee et al}

\begin{document}

\title{Bondi-Hoyle Accretion in an Isothermal Magnetized Plasma}
\author{Aaron T. Lee\altaffilmark{1}, Andrew J. Cunningham\altaffilmark{2}, Christopher F. McKee\altaffilmark{1,3}, Richard I. Klein\altaffilmark{1,2} }
\altaffiltext{1}{Department of Astronomy, University of California Berkeley,
    Berkeley, CA 94720}
\altaffiltext{2}{Lawrence Livermore National Laboratory, P.O. Box 808, L-23, Livermore, CA 94550}   
\altaffiltext{3}{Department of Physics, University of California Berkeley,
    Berkeley, CA 94720}
\email{a.t.lee@berkeley.edu}

\begin{abstract}
 
In regions of star formation, protostars and newborn stars will accrete mass from their natal clouds. These clouds are threaded by magnetic fields with a strength characterized by the plasma $\beta$---the ratio of thermal and magnetic pressures. Observations show that molecular clouds have $\beta\la 1$, so magnetic fields have the potential to play a significant role in the accretion process. We have carried out a numerical study of the effect of large-scale magnetic fields on the rate of accretion onto a uniformly moving point particle from a uniform, non-self-gravitating, isothermal gas. We consider gas moving with sonic Mach numbers of up ${\cal M}\approx 45$, magnetic fields that are either parallel, perpendicular, or oriented $45\degr$ to the flow, and $\beta$ as low as 0.01. Our simulations utilize adaptive mesh refinement in order to obtain high spatial resolution where it is needed; this also allows the boundaries to be far from the accreting object to avoid unphysical effects arising from boundary conditions. Additionally, we show our results are independent of our exact prescription for accreting mass in the sink particle. We give simple expressions for the steady-state accretion rate as a function of $\beta$ and ${\cal M}$ for the parallel and perpendicular orientations. Using typical molecular cloud values of ${\cal M}\sim5$ and $\beta\sim0.04$ from the literature, our fits suggest a $0.4\;M_\odot$ star accretes $\sim 4\times10^{-9}\ M_\odot$/year, almost a factor of two less than accretion rates predicted by hydrodynamic models. This disparity can grow to orders of magnitude for stronger fields and lower Mach numbers. We also discuss the applicability of these accretion rates versus accretion rates expected from gravitational collapse, and under what conditions a steady state is possible. The reduction in the accretion rate in a magnetized medium leads to an increase in the time required to form stars in competitive accretion models, making such models less efficient than predicted by Bondi-Hoyle rates. Our results should find application in numerical codes, enabling accurate subgrid models of sink particles accreting from magnetized media.   
\end{abstract}

\keywords{ISM: magnetic fields --- magnetohydrodynamics (MHD) --- stars: formation}

\section{INTRODUCTION}

Accretion is ubiquitous in astrophysics. With examples including protostellar accretion from molecular clouds, mass transfer between binary companions, and gas falling onto a supermassive black hole in the center of galactic nuclei, understanding how (or whether) a gravitating source gathers mass has received much attention over the past century. In the case of star formation, considerable study has been given to understanding the process of accretion from a background medium. Knowing how much mass a star can accrete from its natal cloud will help elucidate, for example, whether the final mass of the star is determined primarily through gravitational collapse \citep[e.g.,][]{shu} or through post-collapse accretion \citep[e.g.,][]{bonnell97,bonnell01}. Mass accretion also could play a role in the dynamics of stars in clusters. If the accretor is moving relative to the background gas, then the accretion of mass and momentum will be non-spherical, and this may play a role in the radial redistribution of objects in stellar clusters \citep{leestahler11}.

Several physical processes exist for transferring mass from the cloud to the surface of a (proto)star. In core-collapse models \citep{shu}, a dense core's self-gravity induces global gravitational collapse, resulting in supersonic infall either directly onto the stellar surface or into a surrounding centrifugally supported disk. Material that ultimately ends up on the star comes from a local gravitationally bound region of the parent molecular cloud. If the core is not collapsing directly onto the star+disk, or if the core is exhausted and the star is moving through the more tenuous regions of the cloud, another accretion mechanism is at play. Here the local gas initially unbound to the star can be captured and subsequently accreted. The self-gravity of this local gas is negligible relative to the gravity of the star itself. Such accretion is often called Bondi accretion  when the star is stationary or Bondi-Hoyle(-Lyttleton) accretion when the star is moving relative to the background gas, named after the pioneering investigators \citep{hoylelyttleton39,bondihoyle44,bondi52}.

The primary goal of this work is concerned with understanding the steady-state mass accretion rate for Bondi-Hoyle accretion when the background gas is an isothermal plasma pervaded by a magnetic field. In particular, we seek to construct an interpolation formula that  reproduces both known analytic and numerical results as well as the steady-state accretion rates we will obtain via numerical simulations. In our work and these previous works, the effects of stellar winds and outflows are neglected. We begin this study by summarizing some of the known results in the next section. From there, we propose new interpolation formulas for the mass accretion rate of magnetized Bondi-Hoyle flow. This function will have two free parameters, which we fix by fitting to numerical simulations. Section \ref{sec:methods} discusses our methodology and numerical convergence studies of the numerical code. Section \ref{sec:results} presents the numerical results and the results of fitting our proposed interpolation formulas to the simulation data. In Section \ref{sec:sstate} we discuss the applicability of such steady-state models in regions of star formation.  Section \ref{sec:summary} concludes this work with a summary and discussion. How these models can be implemented in sub-grid and sink particle algorithms is discussed in Appendix \ref{sec:sgrid}.

\section{Mass Accretion Rates}

\subsection{Known Results}\label{ssec:massratesknown}

The study of steady-state accretion from an initially uniform background medium has enjoyed many analytical and numerical studies. \citet{edgar04} gives a nice pedagogical review of some of the earlier work. \citet{hoylelyttleton39} first solved the problem for a point particle of mass $M_*$ moving through a collisionless fluid at (hypersonic) speed $v_0$. Matter was focused into a vanishingly thin wake and accreted through a spindle downstream of the accretor. The accretion rate was
\begin{equation}\label{eqn:mdothl}
	\dot{M}_{\rm HL} = 4\pi r^2_{\rm HL} \rho_0 v_0 = \frac{4\pi G^2 M^2_* \rho_0}{v^3_0}\ ,
\end{equation}
for the far-field mass density $\rho_0$. Associated with $v_0$ is the characteristic radius
\begin{equation}\label{eqn:rhl} r_{\rm HL} \equiv \frac{GM_*}{v^2_0}\ , \end{equation}
which measures the dynamic length scale within which gravity wins over the inertia of the gas. In the opposite limit of stationary or subsonic motion, the thermal pressure exceeds the ram pressure of the gas by a factor of $\sim (c_{\rm s}/v_0)^2$ for sound speed $c_{\rm s}$. \citet{bondi52} analytically solved the problem for a stationary accretor, arriving at
\begin{eqnarray}\label{eqn:mdotb}
\dot{M}_{\rm B} = 4\pi\lambda r^2_{\rm B}\, \rho_0 c_{\rm s} = \frac{4\pi\lambda G^2 M^2_* \rho_0}{c^3_{\rm s}}\ ,
\end{eqnarray}
which becomes
\begin{eqnarray}
\label{eqn:mdotb2} = 1.02\times 10^{-6} \left(\frac{M_*}{0.4\ M_\odot}\right)^2\left(\frac{n_0}{10^4\text{ cm}^{-3}}\right)\left(\frac{T}{\text{10 K}}\right)^{-3/2}\frac{M_\odot}{\mbox{yr}}\ .
\end{eqnarray}
Here we have defined the Bondi radius
\begin{equation}\label{eqn:rb}
r_{\rm B} \equiv \frac{GM_*}{c^2_{\rm s}} = 9.0\times 10^{16} \left(\frac{M_*}{0.4\ M_\odot}\right)\left(\frac{T}{\text{10 K}}\right)^{-1}\ \text{cm}\ .
\end{equation}
In our numerical evaluations, we have normalized the temperature $T$ to 10 K, the number density $n_0$ to $10^4$ particles per cm$^3$, and masses to the solar mass $M_\odot$. The mass density is related to the number density by $\rho_0 = (2.34\times10^{-24}\text{ grams})\cdot n_0$. The symbol $\lambda$ is a function of the adiabatic index $\gamma$ ($\lambda=\exp(3/2)/4\approx 1.12$ for an isothermal gas, $\gamma=1$). 

Both limits then established, Bondi proposed his venerable Bondi-Hoyle interpolation formula that connects the stationary and hypersonic regimes:
\begin{equation}\label{eqn:bondiinterp}
	\dot{M}_{\rm BH} = \frac{4\pi \rho_0 r^2_{\rm B} c_{\rm s}}{(1+{\cal M}^2)^{3/2}} = \frac{\dot{M}_{\rm B}/\lambda}{(1+{\cal M}^2)^{3/2}}\ ,
\end{equation} 
where we have introduced the sonic Mach number $\Mach \equiv v_0/\cs$. The characteristic velocity for Bondi-Hoyle accretion is 
\beq\label{eqn:bondiholyevlocity}
	v_{\rm BH}=(c^2_{\rm s} + v^2_0)^{1/2}\ ,
\eeq
and the corresponding Bondi-Hoyle radius is
\begin{equation}\label{eqn:rbh}
	r_{\rm BH} = \frac{GM_*}{v^2_{\rm BH}} = \frac{r_{\rm B}}{1+{\cal M}^2}\ .
\end{equation}
We will see in Section \ref{sec:results} that magnetic fields reduce the accretion rate below these values. Furthermore, for the fiducial values of $n_0$ and $T$ and for $M_* > 0.4 M_\odot$, Bondi accretion is not in a steady state (Section \ref{sec:sstate}).

We shall express all accretion rates in terms of the Bondi accretion rate in two equivalent forms. For example, the Bondi-Hoyle accretion rate will be written as
\beq\label{eqn:mdotbh} \dot{M}_{\rm BH} = \phi_{\rm BH}\cdot 4\pi\lambda r^2_{\rm BH} \rho_0 v_{\rm BH} = \phi_{\rm BH}\left(\frac{c_{\rm s}}{v_{\rm BH}}\right)^3 \dot{M}_{\rm B}\ .
\eeq
This first form emphasizes the underlying physical parameters, and we have introduced  a correction factor $\phi_{\rm BH} = \phi_{\rm BH}({\cal M})$, which will be of order unity. The second form is
\beq\label{eqn:mdotbhinterp} \dot{M}_{\rm BH} = \left(\frac{c_{\rm s}}{v_{\rm BH,eff}}\right)^3 \dot{M}_{\rm B}\ .
\eeq
Here, the effective Bondi-Hoyle velocity $v_{\rm BH,eff}$ is an interpolation formula; the rationale for introducing the second form will become clear below. 

\begin{figure}
\plotone{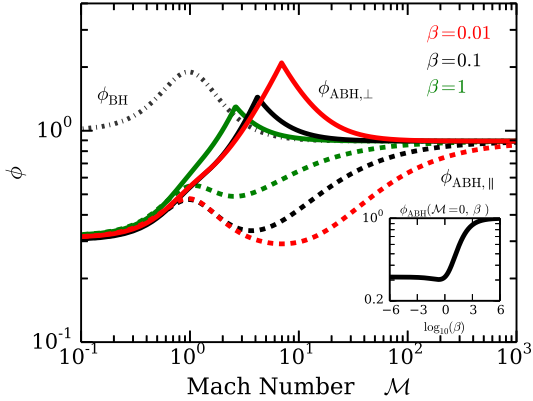}
\caption{Various numerical parameters $\phi$ as a function of Mach number ${\cal M}$ and plasma $\beta$. Colors refer to different magnitudes of $\beta$, whereas the linestyles differ for different $\phi$. The high ${\cal M}$ limit is $1/\lambda$. The subplot shows the stationary limit of $\phi_{\rm ABH}$ as a function of $\beta$. This has the same functional form as $\phi_{\rm AB}$, defined in the text, but with different fitting parameters $\beta_{\rm ch}$ and $n$, which are determined in \S\ref{sec:massaccretion}. The asymptotes are $\sqrt{2/\beta_{\rm ch}}\approx 0.32$ and unity. }
\label{fig:phi}
\end{figure}

Simulations have shown that Bondi's interpolation formula ($\phi_{\rm BH}=1/\lambda$) can be in error by several ten's of percent\ \citep{shimaetal85,ruffert94}. These authors, among others, have considered the non-isothermal case as well and have proposed two-dimensional interpolation formulas (in ${\cal M}$ and $\gamma$) to match simulation results. Typically such formulas are monotonically decreasing functions of both ${\cal M}$ and $\gamma$ and agree well the simulations. A complication is that \citet{ruffert94,ruffert96} has shown that accretion rates do not decrease monotonically as ${\cal M}$ increases, but instead increase near ${\cal M}\sim 1$ and then asymptote to $\dot{M}_{\rm HL}$. For the isothermal case, we have found that
\beq\label{eqn:phiBH}
	\phi_{\rm BH} = \frac{ (1+{\cal M}^2)^{3/2} [ 1+ ({\cal M}/\lambda)^2]^{1/2} }{1+{\cal M}^4}\ ,
\eeq
corresponding to 
\beq\label{eqn:vBHeff}
	{\cal M}_{\rm BH} \equiv \frac{v_{\rm BH,eff}}{c_{\rm s}} = \frac{ (1+{\cal M}^4)^{1/3}}{ [ 1+({\cal M}/\lambda)^2]^{1/6}}\ ,
\eeq
and agrees with the numerical results of \citet[][for $\gamma=1.01$]{ruffert96} and those reported below with a maximum error of 27\%. Observe that $\phi_{\rm BH}\rightarrow 1$ as ${\cal M}\rightarrow0$ and $\phi_{\rm BH}\rightarrow 1/\lambda$ for ${\cal M}\gg 1$. This function is plotted in Figure \ref{fig:phi}. 

Other numerical studies of accretion have studied the role of additional physics like radiation pressure \citep{miloetal09}, turbulence \citep{krumholzetal06}, 
turbulence and magnetic fields in spherically symmetric accretion \citep{shcherbakov08}, the presence of a disk \citep{moeckelthroop09}, or thermal instabilities \citep{gasparietal13}. Our finding is that Equation (\ref{eqn:mdotbh}) with $\phi_{\rm BH}$ given by Equation (\ref{eqn:phiBH}) is a reasonable measure of the accretion rate for isothermal Bondi-Hoyle accretion when additional physics do not play an appreciable role in the dynamics of the gas near the accretor.

One physical effect that could play an important role in the gas dynamics is a global magnetic field. In star-forming regions, there is ample evidence that molecular clouds are pervaded by magnetic fields \citep{crutcher99,mckeeostriker07}, whose strength (i.e., its ability to influence dynamics) can be characterized by the plasma $\beta$, the ratio of the thermal pressure to the magnetic pressure:
\begin{equation}\label{eqn:beta}
	\beta \equiv  \frac{\rho c^2_{\rm s}}{B^2/8\pi} = 2\left(\frac{c_{\rm s}}{v_{\rm A}}\right)^2 = 2\left(\frac{{\cal M}_{\rm A}}{\cal M}\right)^2\ ,
\end{equation}
for magnetic field amplitude $B$. We have also introduced the \Alfven Mach number ${\cal M}_{\rm A}=v_0/v_{\rm A}$, the ratio of the gas velocity to the \Alfven velocity, $v_{\rm A} = B/\sqrt{4\pi\rho}$. Observations of the Zeeman effect, 
linear polarization emission of dust, and the Chandrasekhar-Fermi method \citep[see the review of][]{crutcher2012} have suggested molecular clouds have $\beta$ values of most order unity, but more have $\beta < 0.1$ \citep[e.g.,][]{crutcher99}. 

\citet{cunnetal12}  have studied accretion from a magnetized, isothermal, static medium. For the case in which thermal pressure is negligible (the low-$\beta$ limit) they argued that gas would collapse along the field lines from a distance $\sim r_{\rm B}$ above and below the point mass, and would then accrete from $\sim$ an \Alfven radius, 
\beq\label{eqn:ra} 
r_{\rm A} \equiv \frac{GM_*}{v^2_{\rm A}} = \frac{c^2_{\rm s}}{v^2_{\rm A}}\, r_{\rm B}
\eeq
at velocity $\sim v_{\rm A}$. 
In our notation, 
\beq\label{eqn:mdota1} 
\dot{M}_{\rm A} = \phi_{\rm A}\cdot 4\pi\lambda r_{\rm B} r_{\rm A}\rho_0 v_{\rm A}\ .
\eeq
As a result, the accretion rate varies as $v^{-1}_{\rm A}\propto \beta^{1/2}$,
\beq\label{eqn:mdota2}
	\dot{M}_{\rm A} = \phi_{\rm A}\left(\frac{c_{\rm s}}{v_{\rm A}}\right)\dot{M}_{\rm B} = \phi_{\rm A}\left(\frac{\beta}{2}\right)^{1/2} \dot{M}_{\rm B}\ .
\eeq
\citet{cunnetal12} expressed the accretion rate as
\beq \dot{M}_{\rm A} = \left(\frac{\beta}{\beta_{\rm ch}}\right)^{1/2} \dot{M}_{\rm B}\ ,
\eeq
where $\beta_{\rm ch}$ is a numerical factor; this corresponds to $\phi_{\rm A} = (2/\beta_{\rm ch})^{1/2}$. They estimated $\beta_{\rm ch}\approx 5$, so that $\phi_{\rm A}=0.63$. From here, they generalized this to include a finite temperature (the ``Alfv\'en-Bondi" case). By writing $v_{\rm A},\, r_{\rm A}\rightarrow v_{\rm AB},\, r_{\rm AB}$ in Equation (\ref{eqn:mdota1}), the accretion rate becomes
\begin{eqnarray}\label{eqn:mdotab}
	\dot{M}_{\rm AB} &=& \phi_{\rm AB}\cdot 4\pi\lambda r_{\rm B}r_{\rm AB}\rho_0 v_{\rm AB} = \phi_{\rm AB}\left(\frac{c_{\rm s}}{v_{\rm AB}}\right)\dot{M}_{\rm B}\ , \\
	&=& \left(\frac{c_{\rm s}}{v_{\rm AB,eff}}\right)\dot{M}_{\rm B}\ ,
\end{eqnarray}
where $v_{\rm AB} \equiv (c^2_{\rm s} + v^2_{\rm A})^{1/2}$ and $r_{\rm AB} = GM_* / v^2_{\rm AB}$.\footnote{Note that Equation (16) of \citet{cunnetal12} has a typo, one factor of $r_{\rm AB}$ should be written as $r_{\rm B}$ instead.} The effective Alfv\'en-Bondi velocity, $v_{\rm AB,eff}$, can be chosen to provide an interpolation formula between the \Alfven and Bondi cases that agrees best with the numerical simulations; \citet{cunnetal12} adopted
\beq\label{eqn:vabeff}
	v_{\rm AB,eff} \equiv \left[ c^n_{\rm s} + \left(\frac{\beta_{\rm ch}}{2}\right)^{n/2} v^n_{\rm A}\right]^{1/n} = \left[ 1 + \left(\frac{\beta_{\rm ch}}{\beta}\right)^{n/2}\right]^{1/n} c_{\rm s}\ ,
\eeq
which gives
\beq\label{eqn:mdotab2}
	\dot{M}_{\rm AB} = \left[ 1+ \left(\frac{\beta_{\rm ch}}{\beta}\right)^{n/2}\right]^{-1/n} \dot{M}_{\rm B}\ .
\eeq
They found that $n=0.42$ and $\beta_{\rm ch}=5.0$ gave agreement with their numerical results to within 5\% for $\beta\ge0.01$.

\subsection{Alfv\'en-Bondi-Hoyle Accretion}\label{ssec:fitssection}

We wish to extend the work of \citet{cunnetal12} to the case in which the accreting mass is moving through a magnetized ambient medium. Our primary interest is in star-forming regions, which are approximately isothermal because the dust and the molecules can efficiently radiate the energy supplied by compression; we therefore assume that the gas is isothermal.\footnote{As we discuss in Section \ref{sec:summary}, our results should also be applicable to the central regions of active galactic nuclei.} The magnetic flux in stars is orders of magnitude less than that in the gas from which they form, so most of of the magnetic flux in the accreting gas decouples from the gas and accumulates in the vicinity of the protostar \citep{zhaoetal11}.
As a result, even in cases where the thermal pressure ($\sim\rho c^2_{\rm s}$) or ram pressure ($\sim~\rho v^2$) initially control the dynamics of the gas near the accretor, accretion can redistribute magnetic flux so that the magnetic pressure ($\propto B^2$) eventually dominates the dynamics near the accreting object. 
In steady-state Bondi accretion from a magnetized gas, \citet{cunnetal12} found that even if $\beta$ was initially $> 1$,  
a steady-state was reached when the gas within $\sim r_{\rm AB}$ of the accretor had $\beta \approx 1$. In a steady-state flow where there is relative motion between the gas and the accretor (i.e., magnetized Bondi-Hoyle accretion), we anticipate that 
if ${\cal M}_{\rm A} \ll \text{min}(1,{\cal M})$, the inertia of the gas will play a small role in setting the steady-state accretion rate, so $\dot{M}$ will be well approximated by 
the Alfv\'en-Bondi result (Eq. \ref{eqn:mdotab}). 
If instead ${\cal M}_{\rm A}\gg 1$, the inertia of the gas is able to drag away most of the magnetic flux so that the the magnetic field is not dominant anywhere and
the accretion rate should approach the non-magnetized Bondi-Hoyle limit. We wish to develop an approximate analytic expression for the rate of accretion by a point mass moving at a constant speed through a uniform, isothermal, magnetized medium by further generalizing the above known results. Our expression will therefore also include parameters $n$ and $\beta_{\rm ch}$, which we can then adjust to best reproduce the results of our simulations of magnetized Bondi-Hoyle flow. 

We generalize Equation (\ref{eqn:mdotab}) by replacing $r_{\rm B}$ with $r_{\rm BH}$, $r_{\rm AB}$ with $r_{\rm ABH}$, and $v_{\rm AB}$ with 
\beq\label{eqn:vabhdefine}
 v_{\rm ABH} \equiv (c^2_{\rm s} + v^2_0 + v^2_{\rm A})^{1/2}\ .
\eeq
Here, and throughout the remainder of the paper, $c_s$ is the isothermal sound speed. With
\beq \label{eqn:rabh} r_{\rm ABH} \equiv \frac{GM_*}{v^2_{\rm ABH}}\ ,\eeq
the accretion rate is then
\begin{eqnarray}\label{eqn:mdotabh1}
	\dot{M}_{\rm ABH} &=& \phi_{\rm ABH}\cdot 4\pi\lambda r_{\rm BH} r_{\rm ABH}\rho_0 v_{\rm ABH} \nonumber\\
	&=& \phi_{\rm ABH}\left(\frac{c^3_{\rm s}}{v^2_{\rm BH} v_{\rm ABH}}\right)\dot{M}_{\rm B} \\ \label{eqn:mdotabh2}
	&=& \left(\frac{c^3_{\rm s}}{v^2_{\rm BH,eff} v_{\rm ABH,eff}}\right)\dot{M}_{\rm B} .
\end{eqnarray}

Equations (\ref{eqn:mdotabh1}--\ref{eqn:mdotabh2}) do not take into account the orientation of the flow relative to the ambient magnetic field. In Section \ref{sec:massaccretion} below, we find that we need different interpolation formulas for the cases where the flow is parallel and perpendicular to the magnetic field. For the parallel case, we generalize $v_{\rm AB,eff}$ to $v_{\rm ABH,\parallel,eff}$ with
\beq \label{eqn:vabhpareff}
	\frac{v_{\rm ABH,\parallel,eff}}{c_{\rm s}} = \left[\left(\frac{v_{\rm BH,eff}}{c_{\rm s}}\right)^n + \left(\frac{\beta_{\rm ch}}{\beta}\right)^{n/2}  \right]^{1/n}\ .
\eeq
The accretion rate in this case is
\beq\label{eqn:mdotabhpar}
\dot{M}_{\parallel} = \frac{1}{{\cal M}^2_{\rm BH}} \left\{ {\cal M}^n_{\rm BH} + \left(\frac{\beta_{\rm ch}}{\beta}\right)^{n/2}\right\}^{-1/n} \dot{M}_{\rm B}\ ,
\eeq
where ${\cal M}_{\rm BH}$ was defined in Equation (\ref{eqn:vBHeff}). Observe that this expression reduces to Bondi accretion for ${\cal M} = \beta^{-1}\rightarrow 0$, to Bondi-Hoyle accretion if $\beta\rightarrow\infty$, and to Alfv\'en-Bondi accretion for ${\cal M}=0$ and arbitrary $\beta$. The factor $\phi_{\rm ABH,\parallel}$ in this case is then
\beq\label{eqn:phiabhpar} 
	\phi_{\rm ABH,\parallel} = \phi^{2/3}_{\rm BH}\cdot\left(\frac{v_{\rm ABH}}{v_{\rm ABH,\parallel,eff}}\right)\ ;
\eeq
it is also plotted in Figure \ref{fig:phi}. 

For flows perpendicular to the field, we obtain better agreement with our simulations with the less-intuitive interpolation
\beq\label{eqn:vabhpereff}
	\frac{v_{\rm ABH,\bot,eff}}{c_{\rm s}} \equiv \text{max}\left[\frac{v_{\rm BH,eff}}{c_{\rm s}}, \frac{v_{\rm ABH,\parallel,eff}}{v_{\rm BH,eff}} \right]
\eeq
in Equation (\ref{eqn:mdotabh2}). One can readily verify that this has the correct limits for Bondi, Bondi-Hoyle, and Alfv\'en-Bondi accretion. We then obtain
\beq\label{eqn:mdotabhper}
	\dot{M}_{\bot} = \text{min}\left\{ \frac{1}{{\cal M}^3_{\rm BH}},\frac{1}{{\cal M}_{\rm BH}}\left[ {\cal M}^n_{\rm BH} + \left(\frac{\beta_{\rm ch}}{\beta}\right)^{n/2} \right]^{-1/n}\right\} \dot{M}_{\rm B}.
\eeq
An immediate interesting result of this formulation---that is born out in our simulations, see \S4.2---is for the particular case of highly supersonic flow with an \Alfven Mach number $\ga 1$, the accretion rate for the perpendicular case reduces to
\beq\label{eqn:interestingresult} \dot{M}_\bot = \frac{\dot{M}_{\rm B}/\lambda}{{\cal M}^3} = \dot{M}_{\rm HL}\ \ \ \text{(${\cal M}\gg1,\ {\cal M}_{\rm A}\ge1$)}\ ,
\eeq
even when ${\cal M}_{\rm A}\approx 1$. 

If the point mass is moving through a medium at an angle $\theta$ with respect to the magnetic field, we approximate the accretion rate by
\beq \dot{M} \simeq \dot{M}_\parallel \cos^2\theta + \dot{M}_\bot \sin^2\theta\ .
\eeq
Indeed, we confirm for one of our simulations that when $\theta=45\degr$, the resulting accretion rate is decently approximated by the average of the two limiting rates. If the orientation changes randomly in time, the proposed average accretion rate is
\beq\label{eqn:mdotavg}\dot{M} \simeq \frac{1}{2}\left( \dot{M}_\parallel + \dot{M}_\bot \right)\ .
\eeq

In order to test our proposed interpolation formulas, we study the problem of Bondi-Hoyle accretion in a magnetized plasma using the \Ramses MHD code \citep{teyssier02} over a range of field strengths and sonic Mach numbers relevant for star formation. These simulations employ the adaptive mesh refinement (AMR) capabilities of the code to retain high spatial resolution where it is needed---close to the accreting object---while allowing for a large computational domain to prevent the boundaries of the domain from influencing the steady-state flow. As noted above, we do not consider the effects of stellar winds or outflows on the accretion rate. The numerical methodology is described in the next section.

\section{NUMERICAL METHODS}\label{sec:methods}

Our methods are similar to those described in \citet{cunnetal12}. Here we summarize the methods, highlighting the significant differences in this work, present the results of our convergence study, and point to where the reader can find additional details if interested.

We solve the equations of ideal MHD for an isothermal gas with a fixed point mass at the origin, particularly mass conservation,
\begin{eqnarray}
	\frac{\p\rho}{\p t} + \nabla\cdot \rho\bvv = -S_M(\bx)\ , 
\end{eqnarray}
momentum conservation,
\begin{eqnarray}
	\frac{\p\rho\bvv}{\p t} + \nabla\cdot(\rho\bvv\bvv) = - \nabla\left(  P_{\rm th} + \frac{\bB^2}{8\pi}\right)\nonumber \\  + \frac{\bB\cdot\nabla\bB}{4\pi} -\frac{GM_*\rho}{\bx^2}\hat{\bx}-S_{\rm M}(\bx)\cdot \bvv\ ,
\end{eqnarray}
the induction equation,
\begin{eqnarray}
	\frac{\p B}{\p t} - \nabla\times(\bvv\times\bB) = 0\ ,
\end{eqnarray}
and the equation of state,
\begin{eqnarray}
	P_{\rm th} = \rho c^2_{\rm s}\ .
\end{eqnarray}
Here $\bvv$ is the velocity of the gas, $\bx$ is the position relative to the sink particle, and $\bB$ is the magnetic field. Self-gravity of the gas is neglected. In the code, the point mass is represented by a fixed sink particle located at the center of the computational domain.  The term $S_M$ allows for mass accretion onto the central point mass if gas flows into a sphere of radius $4\Delta x$, where $\Delta x$ is equal to the size of the grid cell on the finest AMR level. The accreted gas's momentum is also removed from the grid, though the particle is held stationary at the center of the domain.\footnote{We note that the absence of the $-S_{\rm M}(\bx)\cdot \bvv$ term in the equations of \citet{cunnetal12} is a typographical mistake.} The sink particle is allowed to accrete mass but not magnetic flux, and it accretes as much mass in a timestep $\Delta t$ as it can without introducing a new local maximum in the \Alfven speed amongst the cells located within a shell with radius $r$ between $4\Delta x$ and $6\Delta x$ from the accreting particle. That is,
\begin{equation}\label{eqn:sink}
S_{\rm M}(\bx) = 
\begin{cases}
	\dis\frac{1}{\Delta t}\; \text{max}\left(0\ , \rho-\dis\frac{B}{4\pi v^2_{\rm A,max}}\right) & \text{if } |\mathbf{x}| < 4\Delta x \\
	0 & \text{otherwise}
\end{cases}\ ,
\end{equation}
where $v_{\rm A,max} = \text{max}(v_{\rm A}(\mathbf{x});\, 4\le|\mathbf{x}|/\Delta x\le6).$ The reader can also see the paragraph containing Equation (7) of \citet{cunnetal12} for more details on the sink particle algorithm.

Since the sink particle accretes mass but not flux, the cells interior to the sink particle radius decouples the gas form the field. In reality, non-ideal MHD effects remove the majority of the accreted gas from the field within the accretion disk $<100$ AU from the star; see equation (48) of \citep{LiMcKee96} or the review of \citet{armitage11}. Our sink particles will typically have a radius of $\sim 500$ AU, so our treatment of non-ideal MHD effects requires a sub-grid model; our prescription was given above.\footnote{Non-ideal effects can also play a role at larger distance ($\lesssim1000$ AU) within shocks that originate from the collision of in-falling gas and the magnetic field that has been freed from accreted material \citep{LiMcKee96}.} Furthermore, this also means that gas just interior to the sink particle radius could be artificially affected by non-ideal effects. Nonetheless, both the exact prescription for how gas is removed from the field lines and the size of the sink particle are unimportant as long as the gas entering the sink region has accelerated to free-fall. If this has occurred, the in-falling gas has causally disconnected from the surrounding medium and any artificial prescriptions cannot alter the far-field gas. We discuss how are results are independent of the sink particle conditions in more detail in Section 4.2. 

In addition to the very small magnetic flux the star gains by accretion, the star might also generate its own field through dynamo action. The fields of newborn stars are observed with strengths of order kGauss, but the dipole component of the field falls off as $(R_*/R)^3$, making the stellar field strength a few $\mu$Gauss at $\sim 10$ AU, which is already smaller than the field in the ISM. Therefore, we neglect the field generated by the star itself.

\begin{figure}
\plotone{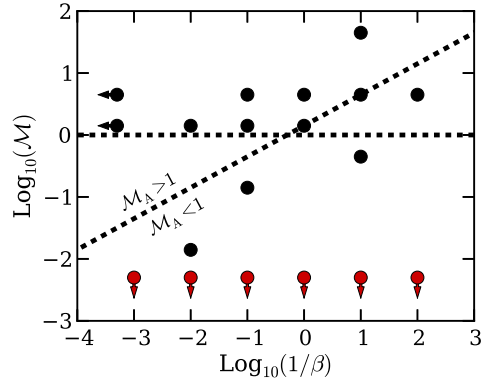}
\caption{Parameter space to be studied. Black dots represent models explored in this work, with the two runs with arrows corresponding to $\b=10^{30}$. Red dots are the stationary models of \citet{cunnetal12}. With this choice of axes, the left vertical axis approximates non-magnetic flow, where the bottom horizontal axis approximates stationary flow. The diagonal line plots ${\cal M}_{\rm A}=1$, while the horizontal line plots ${\cal M}=1$. Our runs explore two regions of this parameter space quite well. The region of ${\cal M}<1$ and ${\cal M}_{\rm A}>1$, not explored by us, was studied by \citet{ruffert94,ruffert96} in his investigations of non-magnetized isothermal Bondi-Hoyle accretion. Since typical star forming regions have ${\cal M}_{\rm A}\approx 1$ \citep{crutcher2012}, we also explore two cases with an Alfv\'en Mach number of unity. 
 }
\label{fig:parameters}
\end{figure}

For all our integrations, the gas is initially uniform with density $\rho_0$ and sound speed $\cs$. The magnetic field is initially set to be uniform in the $\hat{z}-$direction with a magnitude set by $\b$. The speed of the gas is initially set to $v_0$, which is oriented either parallel or perpendicular to the $\bB$-field, except in one case where we orient $\mathbf{v}$ at an angle of 45 degrees. We explore a parameter space of $\b$ and $\Mach$. We consider $\beta$ values of $10^{30},\ 10^2,\ 10,\ 1,\ 0.1,$ and 0.01 and sonic Mach numbers that range from 0.014 to 44.7. For a given $\beta$, we select our velocities to be either equal to the \Alfven velocity or -1, -1/2, 1/2, or 1 decade from this value. This gives us a combination of runs that are both sub and super-sonic as well as sub and super-Alfv\'{e}nic. Table \ref{table:runs} tabulates the parameter space explored, and Figure \ref{fig:parameters} shows this parameter space graphically. The plot identifies four regions of parameter space, depending on whether ${\cal M}$ and ${\cal M}_{\rm A}$ are greater or less than one. Including the stationary runs of \citet{cunnetal12}, our runs explore two of these regions quite well (${\cal M},{\cal M}_{\rm A} >1$ and ${\cal M},{\cal M}_{\rm A}<1$). The empty region (${\cal M} < 1$, ${\cal M}_{\rm A}>1$) is explored by hydrodynamic models of Bondi-Hoyle flow \citep{ruffert96}. The final region (${\cal M}>1$, ${\cal M}_{\rm A}<1$) is only explored by one simulation. Typical star forming regions have ${\cal M}_{\rm A}\approx 1$ \citep{crutcher2012}, so we also explore two cases with an Alfv\'en Mach number of unity. 

\begin{deluxetable*}{cccccccccc}
\tablecaption{Simulation Parameters}
\tabletypesize{\small}
\tablewidth{0pt}
\tablehead{
$\b$ & $\Mach$ & ${\cal M}_{\rm A}$  & $r_{\rm ABH}/\Delta x$  & $t_{\rm end}/t_{\rm B}$ & $\langle {\cal M}_{\rm fast}\rangle^a_{||}$ & $\langle {\cal M}_{\rm fast}\rangle^a_{\bot}$ & $(\dot{M}/\dot{M}_{\rm B})_{||}$ & $(\dot{M}/\dot{M}_{\rm B})_{\bot}$ & $(\dot{M}/\dot{M}_{\rm B})_{45\degr}$ }
\startdata
100 & 0.014  & 0.1  & 161  & 8 & 1.4 & 1.7 & 0.323 & 0.379 &\\
100 & 1.41 & 10  & 54  & 9 & 2.2 & 1.6 & 0.363 & 0.332 &\\
%10 & 0.141  & 0.32  & 134  & 3 & X & X & 0.246 & n/a \\
10 & 1.41 & 3.2  & 51  & 7 & 1.4 & 1.1 &  0.273 & 0.294 &\\
10 & 4.47  & 10  & 8  & 4 & 3.8 & 3.5 & 0.012 & 0.011 &\\
1 & 1.41  & 1  & 33  & 5 & 1.6 & 1.0 & 0.106 & 0.182 & 0.116\\
1 & 4.47  & 3.2  & 7 & 3 & 2.1 & 2.0 & 0.013 & 0.012 &\\
0.1 & 0.447  & 0.1  & 8 & 3 & 1.7 & 1.5 & 0.064 & 0.061 &\\
0.1$^b$ & 4.47  & 1  & 16 & 0.5 & 1.8 & 1.1 & 0.00163 & 0.0112 &\\
0.1$^c$ & 44.7  & 10  & 21 & $3\times10^{-4}$ & 8.2 & 8.6 & $10^{-5}$ & $8.12\times 10^{-6}$ & \\
0.01$^d$ & 4.47  & 0.32  & 2 & 0.5 & 1.3 & 1.00 & 0.0024 & 0.0026 &\\ \hline
$\infty^e$ & 1.41  & n/a  & 55 & 5 & \multicolumn{2}{c}{6.7} & \multicolumn{3}{c}{0.4} \\
$\infty^e$ & 4.47  & n/a  & 8 & 5 & \multicolumn{2}{c}{5.5} & \multicolumn{3}{c}{0.01}\\
\enddata
\tablenotetext{a}{\ Computed as the volume average over the cells 5 and 6 $\Delta x$ from the sink particle. Cells are included if the gas is in-falling (i.e., if $\mathbf{v}\cdot\mathbf{x}<0$.)}
\tablenotetext{b}{\ For this simulation, two additional levels of refinement are allowed for the parallel run, reducing the value of $\Delta x$ by a factor of $2^{2}$. For the perpendicular run, only one additional level is allowed, reducing $\Delta x$ by a factor of 2.}
\tablenotetext{c}{\ For this simulation, one additional level of refinement is allowed and each dimension of the computational domain is reduced from $50\, r_{\rm B}$ to $50/64\, r_{\rm B}$ by a factor of $2^7$, reducing the value of $\Delta x$ by a factor of $2^{7+1}$.}
\tablenotetext{d}{\ For this simulation, one additional level of refinement is allowed, reducing the value of $\Delta x$ by a factor of $2$.}
\tablenotetext{e}{\ For these simulations, $\beta$ is set to $10^{30}$ to approximate non-magnetic flow.}
\label{table:runs}
\end{deluxetable*}
%\end{landscape}

We carried out our computations using the \Ramses code \citep{teyssier02}, an adaptive-mesh-refinement (AMR) code with an oct-tree data structure.
The computational domain is a three-dimensional Cartesian domain with a length of 50\,$\rb$ in each direction. We discretize the domain onto a Cartesian base-level grid of $64^3$. Denote this level as level $L=0$. We allow for seven additional levels of refinement ($L=1,2,...,7$), with each level incrementing the grid-cell density by $2^3$ above the previous level. Each grid cell in the domain is initially refined up to level $L$ if its distance $x$ from the center of the domain satisfies
$$ \left(\frac{25}{2^{L+1}}\right)r_{\rm B}<x < \left(\frac{25}{2^L}\right)r_{\rm B}\ .$$
That is, initially the grid is a set of concentric spheres of increasing refinement as the radius decreases. We also allow for further adaptive refinement if a particular pair of zones has a steep density gradient: if any component of $(\Delta x / \rho)\nabla\rho$ exceeds 1/2, those cells are refined. In this evaluation, the $\rho$ in the denominator is the average of the two cell densities. This second criterion is met only at later times when transient features develop near the sink particle. 

\begin{figure}
\plotone{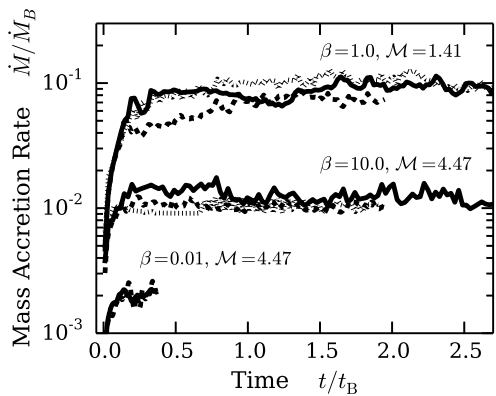}
\caption{Convergence study for three of our marginally resolved models. For the $\beta = 1$ and 10 models, the dotted, solid, and dashed lines represent $L_{\rm max}=6,$ 7, and 8. The solid and dashed represent $L_{\rm max}=8,$ and 9 for the $\beta=0.01$ model. 
%The model shown by the solid curve has seven levels of refinement, whereas the dotted/dashed curves have less or more levels. 
Increasing the number of levels increases the ratio $r_{\rm ABH}/\Delta x$. For all of these runs the velocity and magnetic field are parallel. The sudden jump for $\beta=10,\ L_{\rm max}=6$ at $t/t_{\rm B}\approx 0.7$ occurs because an instability develops in the flow that allows magnetic flux to escape from the region surrounding the sink particle, allowing more mass to accrete. The $\beta=0.01$ case appears converged even though for eight levels of refinement, $r_{\rm ABH}/\Delta x\sim 2$. }
\label{fig:mdotconverge}
\end{figure}

Seven AMR levels sufficiently resolve the relevant lengths scales for the majority of our runs. Since our runs include thermal pressure, gas motion, and magnetic fields, we want to ensure not just that the length scale $r_{\rm AB}$ scale is resolved---as was done in \citet{cunnetal12}---but that the Alfv\'en-Bondi-Hoyle length scale (Equation \ref{eqn:rabh}) is adequately resolved. The maximum level of refinement provides an effective resolution of $\Delta x = 50\,r_{\rm B}/(64\cdot2^7\text{ cells}) = r_{\rm B}/(164\text{ cells})$.  Table \ref{table:runs} tabulates the value $r_{\rm ABH}/\Delta x$. All length scales are resolved by at least 7 cells on the finest level. We note that we are allowing one less level of refinement compared to \citet{cunnetal12}, who allowed up to $L_{\rm max}=8$. Even though some runs have the smallest length scale resolved by $\le 8$ zones, we have confirmed through numerical convergence studies that reducing the number of AMR levels from $L_{\rm max}=8$ to $L_{\rm max}=7$ does not affect the steady-state accretion rate for several of our runs; see Figure \ref{fig:mdotconverge}, where the mass accretion rate for several examples is compared as a function of $L_{\rm max}$. In cases where increasing $L_{\rm max}$ changes the steady state accretion rate by more than 30\%, we include additional levels of refinement until the disparity between simulations diminishes.

The cases $(\b,\Mach)=(0.1,44.7)$ , (0.1,4.47) and (0.01,4.47) require special treatment. For the first, 
$r_{\rm ABH} \approx r_{\rm BH} \approx r_{\rm B}/2000$ is not adequately resolved by even one cell at the finest level of refinement with our standard procedure. Furthermore this implies the sink particle---having a radius of four times the finest grid cell---exceeds the smallest length scale and thus no longer approximates a point particle. Since $r_{\rm BH}$ is orders of magnitude smaller than the other two length scales---and consequently the ratio of pressures $P_{\rm ram}/P_{\rm B}\approx 200$---we expect the inertia of the gas to dominate the dynamics, 
so the time for the flow to reach a steady state should be of 
order $t_{\rm ABH}\approx t_{\rm BH} = r_{\rm BH}/v_0 = t_{\rm B}/\calm^3 \ll  t_{\rm B} = r_{\rm B}/c_{\rm s}$. Unperturbed fluid traverses only a small part of the computational domain in the time required for the flow to achieve steady-state. In order to adequately resolve the Bondi-Hoyle length scale in this case, we reduce the size of the box by a factor of $2^7$, making the length of the domain 
$\approx 780\, r_{\rm BH}$. We also allow one additional level of refinement, making the finest level of refinement smaller by an additional factor of 2
so that $r_{\rm BH}/\Delta x \approx 21$. A steady state is reached in a few $t_{\rm BH}$, so we need not worry about boundary conditions affecting the state of the flow near the accretor.  

For the cases of $(\b,\Mach)=(0.1,4.47)$, and $(0.01,4.47)$, $r_{\rm ABH}/\Delta x\approx 4$ and 1, respectively, when $L_{\rm max}=7$. A convergence study showed that, for the parallel orientation, two additional levels were required for the former case and one for the latter. Convergence is achieved in the latter case even though the ABH length, $r_{\rm ABH}$, is resolved by only about 2 zones. The former required more levels because ${\cal M}_{\rm A}= 1$, and we find that the accretion rates for these cases are most sensitive to the flow morphology $\sim r_{\rm ABH}$ from the accretor, and thus require the most resolution at these scales (see Figure \ref{fig:mdotconverge}). For the perpendicular orientations, only one additional level was required to show convergence. 

All quantities are computed in the cell centers, except for the magnetic field, which is computed on the cell faces. The magnitude of a particular cell's magnetic field is then the average of the magnitude of the cell faces. 

We set $\rho_0 = 10^{-8} (M_*/r^3_{\rm B})$ in all our simulations. The total mass of the gas is then $(50\,r_{\rm B})^3\rho_0 \approx 10^{-3} M_* \ll M_*$, justifying our neglect of the gas's self gravity. Integrations are run to a final time $t_{\rm end}$ sufficiently long to attain a statistically steady accretion rate onto the central particle. Using a stellar mass of $M_\odot$ with $T=10$ Kelvin gas, $r_{\rm B}\sim 22,000$ AU. For our default resolution with seven levels of refinement, the finest level has a resolution of $\sim 135$ AU, and the radius of the sink particle is 540 AU. For the Mach 44.7 run where the box size is reduced, the radius of the sink particle is $\sim2$ AU. For this run only, the stellar field could influence the gas surrounding the sink particle. However, given the high momentum of the gas (${\cal M}_{\rm A}=10$), this run will mimic non-magnetic hydrodynamic flow where additional non-ideal effects play little-to-no role in setting the final accretion rate. 

% =-=-=-=-=-=-=-=-=-=-=-=-=-=-=-=-=-=-=-=-=-=-=-=-=-=-=-=-=-=-
% =-=-=-=-=-=-=-=-=-=-=-=-=-=-=-=-=-=-=-=-=-=-=-=-=-=-=-=-=-=-

\section{RESULTS}\label{sec:results}

\subsection{Morphology}

All of our subsonic runs are  also sub-Alfv\'enic, making the gas morphologies and the final accretion rates well approximated by the stationary models of \citet{cunnetal12}. In this section we describe the supersonic cases, particularly the ${\cal M}=1.41$ and ${\cal M}=4.47$ runs.

\begin{figure}
%\vspace{-0.91in}
\begin{center}\includegraphics[scale=0.27]{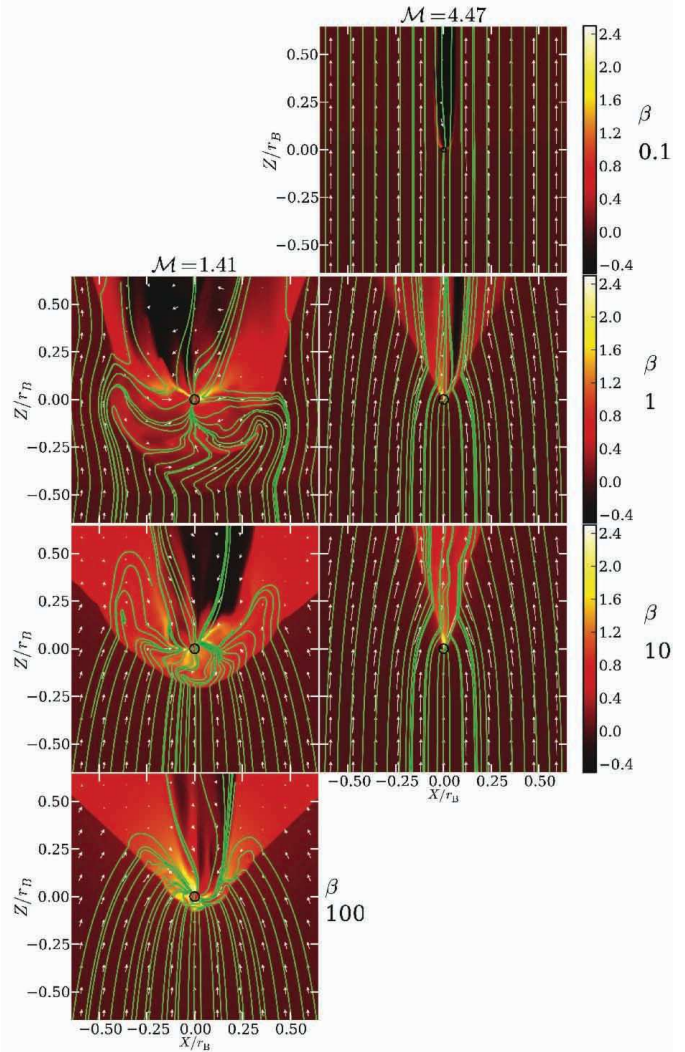}
\end{center}
%\plottwo{Filler.eps}{Bt0.1Pa0LgRho.eps}
%\plottwo{Bt1Pa0LgRho.eps}{Bt1Pa0.5LgRho.eps}
%\plottwo{Bt10Pa0.5LgRho.eps}{Bt10Pa1LgRho.eps}
%\plottwo{Bt100Pa1LgRho.eps}{Filler.eps}
\vspace{-0.25in}\caption{ Slices in the $x-z$ plane showing the region near the sink particle for the parallel orientations. The left and right columns have ${\cal M}=1.41$ and 4.47, respectively. From top to bottom, the rows show $\beta=0.1,1,10$ and 100. All plots are shown at $t=3t_{\rm B}$ except the $\beta=0.1$ plot, which is at $0.5t_{\rm B}$. The color map indicates $\log_{10}(\rho/\rho_0)$, green lines represent magnetic flux tubes drawn from equidistant foot-points 0.5$r_{\rm B}$ upstream of the sink particle, and white arrows represent the flow pattern in the plane of the slice. The black circle indicates the size of the sink particle, equal to $4\Delta x$. Image resolution reduced for ArXiv.}
\label{fig:morphpar}
\end{figure}

\begin{figure}
%\vspace{-0.91in}
\begin{center}\includegraphics[scale=0.27]{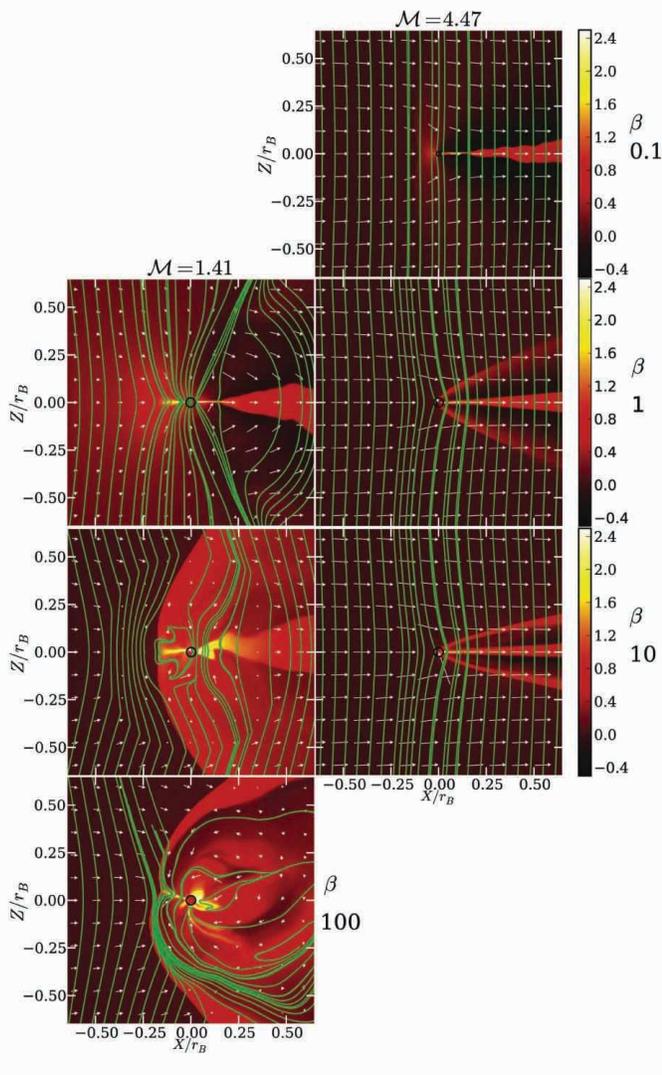}
\end{center}
%\plottwo{Filler.eps}{Bt0.1Pp0LgRho.eps}
%\plottwo{Bt1Pp0LgRho.eps}{Bt1Pp0.5LgRho.eps}
%\plottwo{Bt10Pp0.5LgRho.eps}{Bt10Pp1LgRho.eps}
%\plottwo{Bt100Pp1LgRho.eps}{Filler.eps}
\caption{Same as Figure \ref{fig:morphpar} but for perpendicular orientations. Image resolution reduced for ArXiv.}
\label{fig:morphper}
\end{figure}

\begin{figure}
%\vspace{-1in}
\begin{center}\includegraphics[scale=0.27]{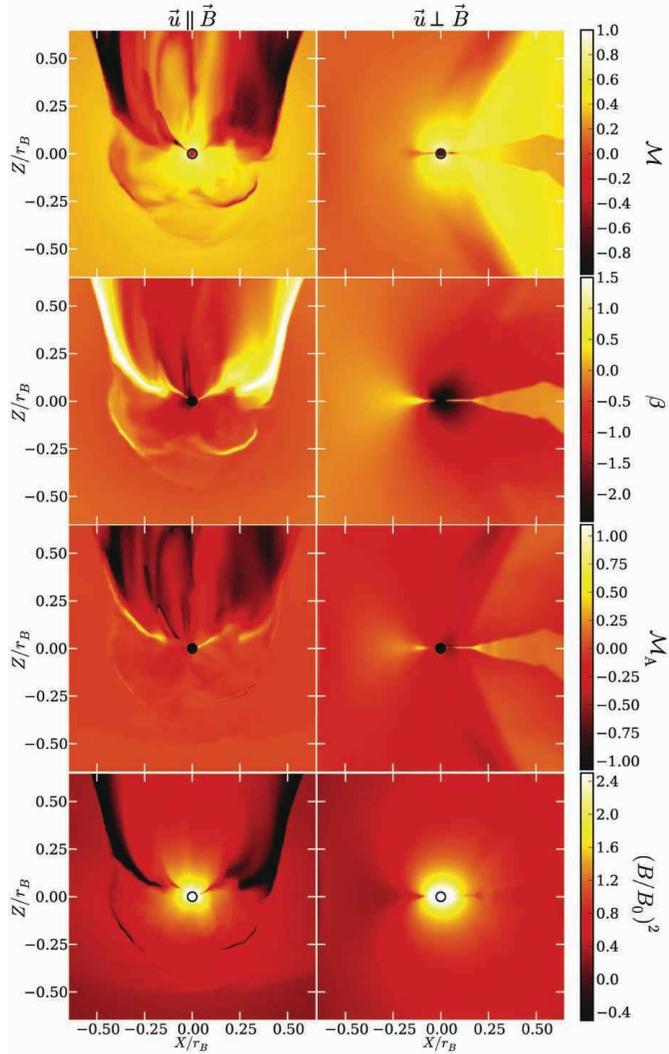}
\end{center}
%\plottwo{Bt1Pa0LgMach}{Bt1Pp0LgMach}
%\plottwo{Bt1Pa0LgBeta}{Bt1Pp0LgBeta}
%\plottwo{Bt1Pa0LgAlfM}{Bt1Pp0LgAlfM}
%\plottwo{Bt1Pa0LgBqSr}{Bt1Pp0LgBqSr}
%\begin{center}\includegraphics[scale=0.35]{figure5}\end{center}
\caption{ Characteristic flow quantities for ${\cal M}=1.41, \beta=1$ at $t=3 t_{\rm B}$. The left and right columns shown the parallel and perpendicular orientations. The top row shows ${\cal M}$, the second row shows $\beta$, the third row shows ${\cal M}_{\rm A}$, and the bottom row shows $(B/B_0)^2$. The colormaps are in $\log_{10}$ space, where the axes are linear (with units of $r_{\rm B}$). The black circle indicates the size of the sink particle, equal to $4\Delta x$. Image resolution reduced for ArXiv. }
\label{fig:flowchar}
\end{figure}

Figures \ref{fig:morphpar} and \ref{fig:morphper} show snapshots late in the simulations after a steady state accretion rate has been established. These two-dimensional slices through the center of the computational domain show the gas density (color bar), velocity of the gas (arrows), and magnetic flux direction (lines) for $\beta\ge0.1$ and ${\cal M}=1.41$ and 4.47. Figure \ref{fig:flowchar} takes the ${\cal M}=1.41$ and $\beta=1$ runs and plots the local values of ${\cal M}$, $\beta$, ${\cal M}_{\rm A}$, and $B^2$ at the same late time. 

The general evolution of the runs goes as follows. Initially, gravity pulls nearby gas towards the sink particle, pinching the magnetic field perpendicular to the far-field flow direction for the parallel orientation, and parallel to the flow for the perpendicular orientation. 
Gas flows relatively undisturbed until it hits the developing shock at the Mach cone or, in some cases, a bowshock propagating upstream. These shocks retard the gas to sub-magnetosonic 
velocities, and the gas continues to flow along field lines downstream of the shocks. Near the source, field lines are drawn towards the sink particle, creating a network of pathways for gas to flow onto the accretor. The extent which field lines can be dragged toward the source depends on the values of ${\cal M}$ and ${\cal M}_{\rm A}$; stronger fields are more resistant to bending (compare, for example, field lines downstream of the shocked region for the ${\cal M}=1.41$ parallel runs in the left panel of Figure \ref{fig:morphpar}). 
Mass-loaded field lines that reach the sink are relieved of their gas, eliminating the gravitational force pulling holding them at the sink. Like a released bowstring, the field snaps back into the surrounding gas (this is prominently shown in the perpendicular orientation for ${\cal M}=1.41$ and $\beta=1$ in Figure \ref{fig:morphper}; here the downstream field lines were recently released). While the mass accretion rate reaches an approximate steady state, 
the morphology of the flow shows larger fluctuations. The details of this morphology depend on the initial orientation of the field to the flow, which we now consider in turn.

\subsubsection{Parallel Orientations}

In the parallel case, there are two types of shock: hydrodynamic, in which $\bB$ is unaffected, and
switch-on shocks, in which a perpendicular component of the field appears behind the shock front
\citep{drainemckee93}. The conditions for the occurrence of a switch-on shock are
(1) $\ma>1$: (2) $\va>c_s$, corresponding to $\beta<2$;
and (3) the post-shock flow must be less than $\va\cos\theta_2$, where the subscript 2 denotes post-shock quantities and $\theta$ is the angle between the magnetic field and the flow velocity. 
The first two requirements ensure that the shock velocity exceeds the fast-wave velocity $v_F$,
which is max$(\va,\; \cs)=\va$ in this case.
The third, post-shock requirement translates to an upper bound on the pre-shock velocity. For isothermal gas ($\gamma=1$), as in our simulations, this upper bound is infinite \citep{drainemckee93}.

\begin{figure}
\plotone{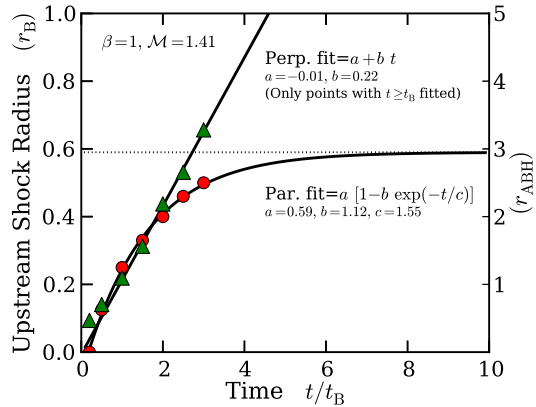}
\caption{Location of the shock front along the upstream axis. To provide data for a fit, the location is plotted at $t/t_{\rm B}=0.2,0.5$, and then every $0.5\, t_{\rm B}$ until $3\, t_{\rm B}$, when the simulations end. The parallel orientation is plotted with circles, perpendicular triangles. A least-squares fit to the function shown is performed. The parallel shock velocity tends to zero at  $\sim 3\, r_{\rm ABH}$, where the flow is unchanged from the background flow. For the perpendicular case, the shock maintains a nearly constant speed equal to the local magnetofast velocity.}
\label{fig:shockfront}
\end{figure}

Gravity amplifies $\rho$ and $B$ relative to the background values $\rho_0$ and $B_0$, but primarily inside $r_{\rm ABH}$. Elsewhere, shocks can also produce density and/or field enhancements. For the parallel orientations, a Mach cone develops immediately for all the supersonic runs and typically extends far beyond $r_{\rm ABH}$. This is the enhanced density region surrounding and downstream from the accretor in Figure \ref{fig:morphpar}. At the later times shown in this Figure, the Mach cone may have joined onto shocks propagating upstream of the accretor, which also may end up disturbing the Mach cone's shape (e.g., the ${\cal M}=1.41$ runs in Figure \ref{fig:morphpar}). In the case of ${\cal M}=4.47$ and $\beta=0.1$, the Mach cone shock front is located only close downstream from the accretor. Here, the unshocked, low-$\beta$, fast-moving gas drags the shocked gas downstream along field lines rather than allowing the Mach cone to extend at the same opening angle. Upstream shocks develops for all ${\cal M}=1.41$ runs, either immediately when $\beta=1\ ({\cal M}_{\rm A}=1)$ or at later times for $\beta>1$. Figure \ref{fig:shockfront} plots the shock front location as a function of time for $\beta=1$ and ${\cal M}=1.41$. A least-squares fit to an exponential function suggests the shock will vanish at $\sim3 r_{\rm ABH}$, where ${\cal M}_{\rm A}$ drops to unity.

As noted above, switch-on shocks occur only for $\beta<2$. Our simulations show that
perpendicular field components can develop in the flow at a finite distance
downstream of the shock; of course, if the shock is not exactly parallel, then the
upstream perpendicular component of the field can be amplified by the compression in the shock. In either case, the perpendicular field component
causes material to pile up on one side of the sink particle, and the inertia of the gas drives kinks in the field farther downstream from the shock (see the left side of Figure \ref{fig:morphpar}). As the field piles up, magnetic pressure eventually dominates and the field straightens itself out, but overshoots, collecting on the opposite side of the accretor. The resulting morphology upstream is an oscillating motion of the field and the flow with a period of $\sim t_{\rm B}$.

Material not immediately upstream of the accretor flows into the Mach cone and then primarily travels through the region immediately downstream of the shock towards the sink particle. However, we note that for these particular simulations, the region of reduced $B$ along the Mach cone is most likely due to numerical reconnection because the field flips orientation over a few grid cells; we have not carried out higher resolution simulations to test if this structure is converged. However, we have ensured that the mass accretion rate is converged, as discussed in the next subsection. The Mach cone results in the downstream being less dense than the background, rather than material forming a downstream wake as in the hydrodynamic limit. 

For ${\cal M}=4.47$, the bowshock only forms for $\beta=0.01$. Even though $\beta=0.1$ gives ${\cal M}_{\rm A}=1$ initially, the region interior to $r_{\rm ABH}$ is so close to the accretor that any decrease with $\beta$ also occurs with an increase in ${\cal M}$, resulting in ${\cal M}_{\rm A}<1$ never being satisfied. For larger $\beta$, even less field enhancement occurs upstream. As a result, the ${\cal M}=4.47$ runs resemble non-magnetic flows, with a downstream ``wake" forming as the region of gas shocked from the Mach cone. The majority of the mass accretion occurs through this wake.

\subsubsection{Perpendicular Orientations}

For the perpendicular orientation, shocks occur if the flow velocity exceeds $v_F$, which
is $(\va^2 +\cs^2)^{1/2}=\cs(1+2/\beta)^{1/2}$ in this case. 
Even if this condition is not initially satisfied, a magnetosonic wave launched from the sink particle boundary can steepen into a weak shock as it moves upstream from the sink particle. We see this, for example, in the ${\cal M}=1.41$, $\beta=1$ case. Initially $v_F=(3/2)^{1/2}v_0$. Immediately upstream of the shock, both $\rho$ and $B$ are increased from compression, but $\rho$ has increased even further from material falling down field lines. This results in an increased $\beta$ and reduced $v_F$. Figure \ref{fig:shockfront} plots this front location. Since $\beta$ has increased to $\sim1.2$, the wave 
(which has steepened into a weak shock)
moves at a speed $\sim(1+2/1.2)^{1/2}c_s-v_0\sim0.22 c_s$ upstream,
relative to the accretor.
Perpendicular to the flow, a weak shock moves outward at $\sim v_F$.

The ${\cal M}=1.41$ runs all show the development of a dense irrotational disk around the accretor interior to $\sim 0.25 r_{\rm B}$, with the disk normal perpendicular to the incoming flow. For $\beta=1$ and 10, this disk attaches to a downstream wake. In the $\beta=100$ case, colliding flows have made the inner $\sim 0.25 r_{\rm B}$ flow unstable, similar to the oscillating flow we saw in the parallel cases. The weak field ends up draped around the shocks that form around the sink particle. 
For ${\cal M}=4.47$, again the flow resembles the non-magnetic case. No bowshock is launched except in the $\beta=0.1$ case, but, as discussed above, this is a transient of the flow. 

We also perform one run at $\beta=1$, ${\cal M}=1.41$ with a 45-degree angle between the flow and magnetic field. Within $\sim r_{\rm ABH}$, the flow tends to align itself with the local magnetic field, and the general flow resembles that of the parallel orientation. The remaining runs, with ${\cal M}=44.7$ and $\b=0.1$ or with $\Mach=1.41,\ 4.47$ and $\b=10^{30}$, are dynamically dominated by the ram pressure of the gas and therefore closely resemble non-magnetic hydrodynamic flow.

\subsection{Mass Accretion Rate}
\label{sec:massaccretion}

For each simulation, the mass accretion rate, $\dot M$, rises with time and then levels off after a few $t_{\rm B} = r_{\rm B}/c_{\rm s}$. Each simulation is run until the rate plateaus for a least a few $t_{\rm ABH}$. The rate quoted in Table 1 is the average over the last 1/6 of the integration, following \citet{cunnetal12}. 

\begin{figure}
\plotone{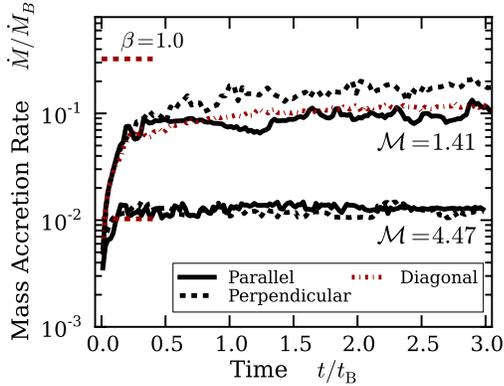}
\caption{Mass accretion rates as a function of time and field orientation for $\beta=1.0$. All rates are normalized to the Bondi accretion rate (Equation \ref{eqn:mdotb}). The short (red) lines identify the steady-state Bondi-Hoyle accretion rates (Equation \ref{eqn:mdotbh}). }
\label{fig:ssexam}
\end{figure}

Figure \ref{fig:ssexam} gives an example of the time evolution of $\dot{M}$ for the $\beta=1$ runs. The rate is averaged over $0.2t_{\rm B}$ bins to reduce noise. In general, when ${\cal M}<1$ or ${\cal M}_{\rm A}\gg1$, the orientation of the field makes little difference in the final accretion rate despite the very different morphologies. When ${\cal M}_{\rm A}\approx 1$, the perpendicular rate always exceeds the parallel rate. When ${\cal M}\gg1$, $\dot{M}_{\rm HL}$ well approximates the perpendicular accretion rate, even, surprisingly, when ${\cal M}_{\rm A}\approx 1$. 

In the high $\beta$ ($\ge 100$) subsonic runs of both our work and that of \citet{cunnetal12}, the accretor undergoes a period of rapid accretion before the accretion rate suddenly drops to $\sim 1/2$ of the original value \citep[see Figure 6, right panel of][]{cunnetal12}. The reason for this effect is that the magnetic field eventually becomes dynamically dominant after enough flux has built up near the accretor. \citet{cunnetal12} showed that this occurs after $t \ga (\beta/100)^{1/2} t_{\rm B}$ for a dynamically weak field and could take an arbitrarily long time as $\beta\rightarrow\infty$. We note that we do not see this effect for our ${\cal M}=1.41$ run with $\beta=100$. 
In this case, the magnetic flux is unable to appreciably build up within $r_{\rm ABH}\sim r_{\rm B}/4$ before the gas pulls the flux downstream. Below, when we determine the best fit parameters for our interpolation formulas, we use the initial (larger) steady-state value for the mass accretion rate
since in astrophysical applications the flow is often not steady for long time periods.

Two requirements are needed to ensure that the accretion rate has converged. First, the value of ${\dot M}$ should not depend on the resolution of the grid. As explained in \S3, we have verified that this is the case. Increasing the resolution also decreases the size of our sink particle, which has a radius of $4\Delta x$. The second requirement is to ensure that the sink particle boundary conditions cannot influence the value of $\dot{M}$. To do this, we require that the accreting gas pass through the fast magnetosonic point at $r > 4 \Delta x$, so that it becomes causally disconnected from the ambient medium before encountering the sink region. For each run, we calculate the mass-weighted volume average of ${\cal M}_{F}=v/v_F$ for the cells either 5 or 6 $\Delta x$ from the sink particle. 
Recall that gas is removed from flux tubes inside the sink region. The resulting low-density flux tubes
are interchange unstable and will rise away from the accretor. Since we are interested in verifying that
the accreting gas is causally disconnected from the ambient medium, 
we include only accreting gas (i.e., gas from cells where $\mathbf{v}\cdot\mathbf{x}<0$, where $\mathbf{x}$ is the position vector from the sink particle's center) in calculating the average of $\calm_F$. 
This averaged value is given in Table 1. For all our runs, we confirm that we have captured the transition.

With the simulation data, we can now determine $\beta_{\rm ch}$ and $n$ in our proposed interpolation formulas (Equations \ref{eqn:mdotabhpar} and \ref{eqn:mdotabhper}) from Section \ref{ssec:fitssection} . Recall that these formulas generalized and built upon previous known analytic and numerical results. In particular, we have proposed a simple interpolation formula for the Bondi-Hoyle limit (Equation \ref{eqn:mdotbh}), which matches the simulation results of \citet{ruffert96}. We performed two $\beta=10^{30}$ runs to verify this equation, finding it underestimates the true accretion rate by only 19\% and 3\% for ${\cal M}=1.41,\ 4.47$, respectively. The mass accretion rate for the diagonal case lies between the the parallel and perpendicular rates; the  average of the predicted parallel and perpendicular rates (Equation \ref{eqn:mdotavg}) reproduces this diagonal case to 18\%.

We perform a least-squares fit to Equations (\ref{eqn:mdotabhpar}) and (\ref{eqn:mdotabhper}) using the union of data from this work and from \citet{cunnetal12}. Since the values of the accretion rate can vary over orders of magnitude, we define the residuals in the least-squares function to be the difference of the logarithms rather than of the absolute values:
\beq
	S = \sum ( \log_{10}\dot{M}_{\rm data} - \log_{10}\dot{M}_{\rm fit})^2\ .
\eeq
Each data point is given an equal statistical weighting. Minimizing $S$, we find $\beta_{\rm ch}=18.3\pm 0.004$ and $n=0.94\pm 0.15$ with $S=0.956$. We do not include the diagonal run or the two hydro runs in our fit. 

The standard errors show that matching the data to these interpolation formulas is not terribly sensitive to the exact value of $n$. Since the data are consistent with using $n=1$, we adopt this value for simplicity. Fixing $n=1$, a least-squares fit to the data yields $\beta_{\rm ch} = 19.8\pm 0.006$ with $S=0.96$. We fix $\beta_{\rm ch}$ as $19.8$. 

\begin{figure}
\plotone{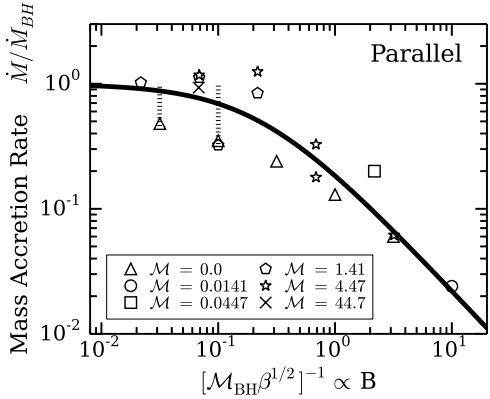}
\caption{Steady-state mass accretion rates for the parallel orientations as a function of the plasma beta $\beta$ (horizontal axis) and sonic Mach number $\Mach$ (symbols). Here we have defined ${\cal M}_{\rm BH} = v_{\rm BH,eff}/c_{\rm s}$. All accretion rates are normalized to our Bondi-Hoyle accretion rate (Equation \ref{eqn:mdotbh}). The solid line is our best fit $\dot{M}_{\parallel}$ with $\beta_{\rm ch}=19.8$ and $n=1.0$ (Equation \ref{eqn:mdotabhpar} or \ref{eqn:mdotparfitbh}). Subsonic runs with $\beta\ge100$ are plotted with their second steady-state value. The dashed lines connect these points to their initial steady state values (no data point shown). }
\label{fig:mdotpar}
\end{figure}

\begin{figure}
\plotone{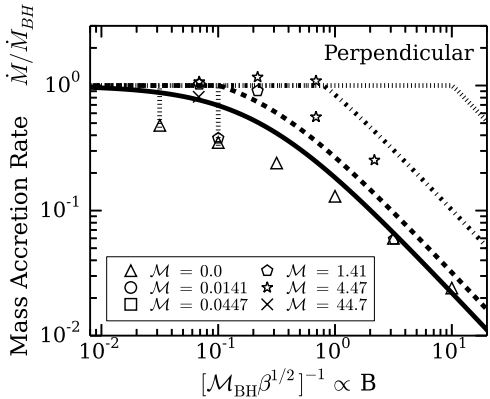}
\caption{Steady-state mass accretion rates for the perpendicular orientations as a function of the plasma beta $\beta$ (horizontal axis) and sonic Mach number $\Mach$ (symbols). All accretion rates are normalized to our Bondi-Hoyle accretion rate (Equation \ref{eqn:mdotbh}). The lines are our best fit $\dot{M}_{\bot}$ with $\beta_{\rm ch}=19.8$ and $n=1.0$ (Equation \ref{eqn:mdotabhper} or \ref{eqn:mdotperpfitbh}) for four values of ${\cal M}$: solid ${\cal M}=0$; dashed ${\cal M}=1.41$; dot-dashed ${\cal M}=4.47$; and dotted ${\cal M}=44.7$. Subsonic runs with $\beta\ge100$ are plotted with their second steady state value. The dashed lines connect these points to their initial steady state values (no data point shown).}
\label{fig:mdotperp}
\end{figure}

In Section \ref{ssec:fitssection} we wrote all accretion rates in terms of $\dot{M}_{\rm B}$, which is constant across our entire parameter space. Since we are studying gas initially in uniform motion, normalizing to $\dot{M}_{\rm BH}$ (Equation \ref{eqn:mdotbhinterp}) is also useful. Indeed with this normalization, our parallel accretion rate can be written in terms of one parameter
\begin{equation}
	{\cal M}_{\rm BH}\beta^{1/2} =2^{1/2} {\cal M}_{\rm BH} \left(\frac{c_s}{\va}\right)\ ,
\end{equation}
which varies as $B^{-1}$ if the other parameters are held constant (${\cal M}_{\rm BH}$ is defined in Equation \ref{eqn:vBHeff}; recall that ${\cal M}_{\rm BH}\rightarrow1$ as ${\cal M}\rightarrow0$). Equations (\ref{eqn:mdotabhpar}) and (\ref{eqn:mdotabhper}) can be rewritten as
\begin{eqnarray}\label{eqn:mdotparfitbh}
	\frac{\dot{M}_\parallel}{\dot{M}_{\rm BH}} &=& \left[ 1 + \frac{4.4}{({\cal M}_{\rm BH}\beta^{1/2})}  \right]^{-1} \ , \\
	\label{eqn:mdotperpfitbh} \frac{\dot{M}_\perp}{\dot{M}_{\rm BH}} &=& \text{min}\left\{1\ ,\ {\cal M}_{\rm BH}\left[1+\frac{4.4}{({\cal M}_{\rm BH}\beta^{1/2})}\right]^{-1}\right\}\ .
\end{eqnarray}
The perpendicular rate, however, potentially requires knowledge of both ${\cal M}_{\rm BH}$ and $\beta$ individually. Figures \ref{fig:mdotpar} and \ref{fig:mdotperp} plot these fits for the parallel and perpendicular orientations. These fits are able to reproduce the simulation data to within a factor of three. 

\begin{figure}
\plotone{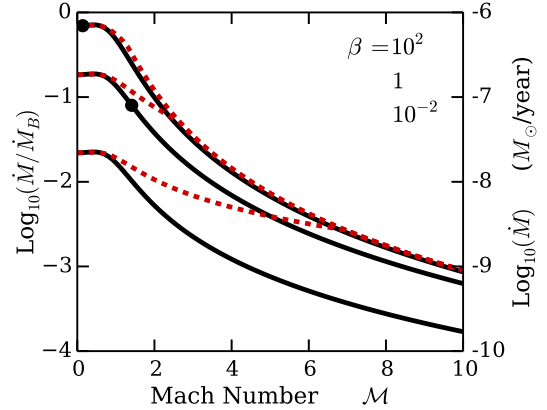}
\caption{Mass accretion rate as a function of sonic Mach number (x-axis) and plasma $\beta$. From top to bottom, the curves represent decreasing $\beta$ values. The solid curves show the parallel fit while the dashed curves show the perpendicular fit (Equations \ref{eqn:mdotabhpar} and \ref{eqn:mdotabhper}). The right y-axis uses the fiducial parameters given in Equation (\ref{eqn:mdotb}) $M_*=0.4\,M_\odot$~yr\e, $n_0=10^4$~cm\eee\ and $T=10$~K. The points identify where ${\cal M}_{\rm A}=1$. The fits are identical at low Mach numbers, and the perpendicular rate always equals or exceeds the parallel rate. Once ${\cal M}_{\rm BH,eff} > {\cal M}_{\rm ABH,eff}/{\cal M}_{\rm BH,eff}$ (Equation \ref{eqn:mdotabhper}), the perpendicular fit becomes identical to the hydrodynamic fit, which is well-approximated by the $\beta=100$ curves.  }
\label{fig:mdotvalues}
\end{figure}

For typical molecular cloud values of ${\cal M}\sim 5$ and $\beta\sim 0.04$ \citep{crutcher99}, which corresponds to ${\cal M}_{\rm A}=0.71$, ${\cal M}_{\rm BH}=5.15$, and ${\cal M}_{\rm BH}\beta^{1/2}=1.03$, Equation (\ref{eqn:mdotavg}) gives an accretion rate of $4.2\times10^{-3}\ \dot{M}_{\rm B}$. For the fiducial parameters given in Equation (\ref{eqn:mdotb2}), this corresponds to $4.3\times10^{-9}\ M_{\odot}$~yr\e, which is not that different from the hydrodynamical prediction (Equation \ref{eqn:mdotbhinterp}) of $\dot M_{\rm BH}= 6.9\times 10^{-9}\ M_{\odot}$~yr\e. However, for smaller Mach numbers of $\sim 2$ and 0, the ratio of the predicted average accretion rate to the hydrodynamic value decreases to 0.19 and 0.048, respectively. Figure \ref{fig:mdotvalues} plots our parallel and perpendicular fits, normalizing to $\dot{M}_{\rm B}$ (Equation \ref{eqn:mdotb}). Here the disparity between the perpendicular and parallel fits can be seen, especially when ${\cal M}_{\rm A}<1$. We remind the reader that our runs only have one instance where $\ma<1$ and ${\cal M}>1$: $\beta=0.01$ and ${\cal M}=4.47$, and we found little difference between the parallel and perpendicular rates, whereas the fits predict the perpendicular rate should be 4.6 times the parallel rate. However, this region of parameter space is only a small region of the overall parameter space (Figure \ref{fig:parameters}) and our fits do predict the disparity between the parallel and perpendicular rates in the other three regions, as well as when ${\cal M}_{\rm A}=1$. Since we chose our non-magnetized limit to well reproduce the results of \citet{ruffert96}, our fits also succeed in predicting the accretion rates for ${\cal M}<1$ and ${\cal M}_{\rm A}>1$, even though we performed no runs ourselves in this region of parameter space.

% =-=-=-=-=-=-=-=-=-=-=-=-=-=-=-=-=-=-=-=-=-=-=-=-=-=-=-=-=-=-
% =-=-=-=-=-=-=-=-=-=-=-=-=-=-=-=-=-=-=-=-=-=-=-=-=-=-=-=-=-=-
\section{VALIDITY OF THE STEADY-STATE APPROXIMATION}\label{sec:sstate}

In both this work and that of \citet{cunnetal12}, we have made several approximations in our analysis of Bondi- and Bondi-Hoyle-type accretion: (1) the accretion must be in a steady-state
(at least when averaged over times $\sim \tb=\rb/c_{\rm s}$); 
(2) the accreting gas must not be self-gravitating; and (3) the accretion rate must be determined by the mass of the particle, not by the gravitational collapse of the ambient medium. As we shall see, these approximations are all connected. We have also assumed that the ambient medium is uniform and that the particle is small compared to $r_{\rm ABH}$, but we shall not
discuss these approximations here. To keep our discussion simple, we restrict ourselves to Bondi accretion.

We define the Bondi mass as 
\beq\label{eqn:bondimass} 
\Mb \equiv 4\pi\rho_0 r^3_{\rm B}\ ,
\eeq
so that the Bondi accretion rate is
\beq
\mdotb \simeq \frac{\Mb}{t_{\rm B}}\ ,
\eeq
where the approximation consists of setting $\lambda\simeq 1$. The Bondi mass is the mass of gas located within the Bondi radius of the particle and is approximately the mass accreted within one Bondi time. For steady-state accretion, the mass of the particle must change slowly, i.e., the mass accreted in
one Bondi time must be small compared to the particle mass: $\mdotb t_{\rm B} \ll M_*$. Therefore, the steady-state approximation reads
\beq\label{eqn:sstateassumption}
\frac{\Mb}{M_*}\simeq\frac{\mdotb t_{\rm B}}{M_*}\ll 1\ .
\eeq

The self-gravity of the ambient gas is characterized by the gravitational mass,
\beq\label{eqn:gravmass} 
\Mg \equiv \frac{c^3_{\rm s}}{\sqrt{G^3 \rho_0}}\ ;
\eeq
the Bonnor-Ebert mass, the maximum mass of an isothermal sphere in hydrostatic equilibrium, is $1.182 \Mg$ \citep{ebert55,bonnor56}. We then have the identity
\beq\label{eqn:mbmg}
\Mb M^2_{\rm G} = 4\pi M^3_*\;,
\eeq
which implies
\beq\label{eqn:mbmgratio}
\frac{\Mb}{\Mg}=\frac{1}{\sqrt{4\pi}}\left(\frac{\Mb}{M_*}\right)^{3/2} \ll 1
\eeq
(Equation \ref{eqn:sstateassumption}). The condition for the accreting gas to be non-self-gravitating is that the mass inside the Bondi radius be small compared to the gravitational mass, $\Mb\ll \Mg$. Equation (\ref{eqn:mbmgratio}) then implies that {\it Bondi accretion is in a steady state if and only if the accreting gas is not self gravitating}.\footnote{\citet{leestahler11} also showed that a steady state is realizable for Bondi-Hoyle accretion when the gas is not self-gravitating.} In other words, the first two approximations listed at the start of this section are really only one approximation.
The steady-state approximation, together with the identity (\ref{eqn:mbmg}), place an upper bound on the particle mass,
\beq
\sqrt{4\pi} \;\frac{M_*}{\Mg}=\left(\frac{\Mb}{M_*}\right)^{1/2}\ll 1\ ,
\eeq
so that
\beq\label{eqn:maxmass}
M_*\ll 0.36\left(\frac{T}{10\,\mbox{K}}\right)^{3/2}\left(\frac{10^4\;\mbox{cm\eee}}{n_0}\right)^{1/2}~~~~M_\odot\ ,
\eeq
where $n_0$ is the density of hydrogen nuclei in the ambient gas. 
The right-hand side of Equation (\ref{eqn:maxmass}) is just $\Mg/\surd (4\pi)$. If $M_* \ga \Mg/\surd (4\pi)$, 
the gas mass within $\rb$ is massive enough to be self-gravitating, and therefore the mass of the particle will not change slowly. For example, if $M_* = 1.0\,M_\odot$, then for the fiducial parameters above, $M_{\rm B} \approx 2.5\ M_\odot$, which exceeds $M_{\rm G} \approx 1.3\ M_\odot$:
The fact that $\Mb$ exceeds $\Mg$ means that the gas is self-gravitating, and the fact that $\Mb$ exceeds $M_*$ means that that accretion is not in a steady state. Note that Equation (\ref{eqn:maxmass}) is based on the assumption that turbulence inside $\rb$ is negligible; if turbulence is important on that scale, as it may be in regions of high-mass star formation, then the theory presented here would have to be generalized, as it was for the non-magnetic case by
\citet{krumholzetal06}.

Finally, we compare the accretion rate due to gravitational collapse,
\beq\label{eqn:gravmassdot}
\dot{M}_{\rm G} \sim \frac{c^3_{\rm s}}{G}\ ,
\eeq
\citep{shu} with that due to Bondi accretion. Observe that the mass accreted due to gravitational collapse in one Bondi time is
very large, $\dot\Mg t_{\rm B}\sim M_*$, so that
\beq
\frac{\dot{M}_{\rm G}}{\mdotb}\sim \frac{M_*}{\Mb}\gg 1\; ,
\label{eqn:mdgmdb}
\eeq
for steady Bondi accretion.
We therefore have the apparently paradoxical result that the rate of accretion via gravitational collapse greatly exceeds that due to Bondi accretion when the latter is in steady state ($\Mb\ll M_*$). 
There are then two possibilities for normal Bondi accretion:
First, the cloud in which the accreting particle is embedded could be gravitationally stable, making $\dot{M}_{\rm G}$ not meaningful (for the simple isothermal,
unmagnetized case we are considering, that requires that the cloud mass, $\Mc$, be less than $\Mg$). Second, 
gravitational collapse could occur on a large scale, but not be focused on the accreting particle. A real
molecular cloud is turbulent and inhomogeneous, and it can undergo gravitational contraction without
having mass accumulate at a central point. Our analysis is valid provided the cloud is approximately uniform
within a Bondi radius of the accreting particle.

\begin{figure}
\plotone{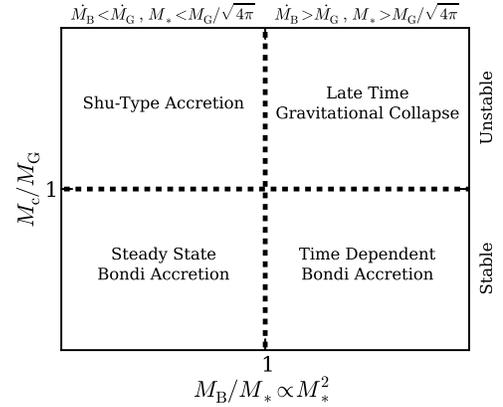}
\caption{Method of accretion as a function of the stellar mass $M_*$ and the host core's mass $M_{\rm c}$. The core is stable against its own self gravity if its mass is less than $M_{\rm G}$ (Equation \ref{eqn:gravmass}). The gas interior to radius $r_{\rm B}$ is not self gravitating if this gas mass $M_{\rm B}$ is less than $M_*$ (or equivalently, $M_* < M_{\rm G}/\sqrt{4\pi}$). If both of these relations are satisfied, steady-state Bondi accretion occurs. Once the stellar mass grows so that $M_{\rm B} \ga M_*$, the accreted mass is comparable to $M_*$ and this accretion rate becomes time dependent. Regardless, if the core is collapsing via its own self-gravity, accretion occurs by gravitational collapse and a steady state is never realized. }
\label{fig:accretiontypes}
\end{figure}

In summary, Bondi-type accretion is in steady state if and only if the gas inside the Bondi radius is not self-gravitating,
\begin{equation}\label{eqn:steadystaterelation}
	\text{Steady state } \leftrightarrow\ \Mb \ll M_* \ll M_{\rm G}\ .
\end{equation} 
Figure \ref{fig:accretiontypes} shows this schematically. The steady-state condition places an upper limit on the particle mass, $M_*\ll\Mg/\surd (4\pi)$ (see Equation \ref{eqn:maxmass}).

This discussion is directly relevant to the issue of whether stars form by gravitational collapse or competitive accretion
\citep{bonnell01,krumholzetal08}. The isothermal sound speed, $c_{\rm s}$, 
must be replaced by the one-dimensional velocity dispersion,
$\sigma$, in both $\Mg$ \citep{mckeeostriker07} and $\Mb$ \citep{krumholzetal06}. In order for competitive accretion to dominate, one requires $\dot\Mb\gg\dot \Mg$, and according to Equation (\ref{eqn:mdgmdb}) this implies $\Mb\gg M_*$: The accreting gas must be self-gravitating, the accretion is not in a steady state, and the upper limit on the protostellar mass in Equation (\ref{eqn:maxmass}) does not apply. However, Equation (\ref{eqn:mbmg}) then implies that $\Mg\ll M_*$. Since
$M_*$ is less than the mass of the cloud from which it is accreting, $\Mc$, it follows that $\Mg\ll\Mc$: the cloud is very sub-virial. That is, the ratio of the kinetic energy in the cloud to the gravitational energy, which is of order the virial parameter  $\avir\equiv 5\sigma^2R/G\Mc$, is much less than unity. \citet{krumholzetal08} argued that since molecular clouds appear to have virial parameters of order unity, this implies that stars form by gravitational collapse, not competitive accretion. If the observed virial parameter reflects collapse instead of turbulence, then competitive accretion may be viable, but the accretion rates would be reduced by magnetic fields.

% =-=-=-=-=-=-=-=-=-=-=-=-=-=-=-=-=-=-=-=-=-=-=-=-=-=-=-=-=-=-
% =-=-=-=-=-=-=-=-=-=-=-=-=-=-=-=-=-=-=-=-=-=-=-=-=-=-=-=-=-=-
%=-=-=-=-=-=-=-=-=-
%=-=-=-=-=-=-=-=-=-

\section{SUMMARY AND DISCUSSION}\label{sec:summary}

The accretion of gas onto an object due to its gravity is generally referred to as Bondi accretion when the object is stationary and Bondi-Hoyle accretion when the object is moving. Such accretion has been employed in many observational, theoretical, and numerical studies to explain the growth of planets, brown dwarfs, stars, compact objects, and  supermassive black holes, to name a few \citep[e.g.,][]{hopkinsetal06,KokuboIda12,andreetal12,toropina12}. Here we have determined the effects of a uniform magnetic field on Bondi-Hoyle accretion under the assumption that the gas is isothermal and that the accreting mass is a point particle, thereby generalizing the results of \citet{cunnetal12} to include the effects of the motion of the accretor. In keeping with most previous treatments of Bondi and Bondi-Hoyle accretion, we did not consider the effects of stellar winds and outflows on the steady-state accretion rate.

Our primary application is to protostellar accretion, 
but our results should apply to stellar accretion in any medium in which the gas is approximately isothermal. We have not considered the effects of stellar winds, which in some cases are strong enough to suppress accretion. Our results might also be applicable to accretion onto supermassive black holes in active galactic nuclei. There the Bondi radius is $\rb\simeq 3(M/10^8 M_\odot)(10^7\,\mbox{K}/T)$~pc. Compton heating and cooling can maintain isothermality near the Bondi radius if the luminosity is sufficiently high \citep{woods96}. In some cases, Compton-heated gas is thermally unstable, and \citet{gasparietal13} have shown that then Bondi accretion rates based on the temperature of the hot gas can underestimate the true accretion rate by up to two orders of magnitude.

The time-averaged mass-accretion rate for isothermal accretion flow onto a static point mass of mass $M_*$ was determined by \citet{bondi52}, $\mdotb=4\pi\lambda\rho_0\rb^2c_s$, where $\lambda$ is a numerical constant, $\rho_0$ is the ambient density, $\rb=GM_*/c_s^2$ is the Bondi radius, and $c_s$ is the isothermal sound speed. If the object is moving, then the morphology of the accretion flow and the accretion rate also depend on the sonic Mach number, ${\cal M}=v_0/c_{\rm s}$, where $v_0$ is the velocity of the mass through the ambient medium. If the medium is magnetized, two additional parameters enter, the plasma $\beta=8\pi \rho_0 c_{\rm s}^2/B_0^2$ and $\theta$, the angle between the the field and the velocity of the object relative to the medium. (For moving objects, $\beta$ can be replaced by the \Alfven Mach number, ${\cal M}_{\rm A}=v_0/\vao=(\beta/2)^{1/2}\calm$, where $\vao$ is the Alfv\'en velocity in the ambient medium). When both ${\cal M}$ and ${\cal M}_{\rm A}$ are large, the accretion resembles non-magnetized Bondi-Hoyle flow. When either ${\cal M}$ or ${\cal M}_{\rm A}$ is small, the ambient medium is approximately static and the flow resembles the stationary magnetized models of \citet{cunnetal12}. Here we have explored the case in which both magnetic fields and motion of the mass through the medium are important by performing three-dimensional simulations of a gravitating point particle accreting from an initially uniform, isothermal gas pervaded by a uniform magnetic field that is either parallel or perpendicular to the direction of motion. Since the magnetic flux in stars is small compared to that in the gas from which the stars formed, we assume that only gas, not magnetic flux, accretes onto the point mass (discussed in \S\ref{sec:methods}). Our main results are approximate expressions (\ref{eqn:mdotabhpar}) and (\ref{eqn:mdotabhper}) for the accretion rates, which reduce to known numerical and analytic limits and agree with our simulation data and that of \citet{cunnetal12} to within a factor of three (see Figures \ref{fig:mdotpar} and \ref{fig:mdotperp}).

The key assumption underlying the theory of Bondi-Hoyle accretion is that the gas is not self-gravitating on the scale of the Bondi radius or, equivalently (as shown in Section \ref{sec:sstate}) that the accretion rate is steady after averaging over
the fluctuations that occur on time scales $\la \tb$.
This assumption must be validated for each astrophysical situation that employs it. The conditions for the validity of the steady-state assumption are that the stellar mass be larger than the Bondi mass, $\Mb=4\pi\rho_0\rb^3$, but smaller than $\Mg$, the mass at which self-gravity becomes important, so that $\Mb\ll M_*\ll\Mg$ (Equation \ref{eqn:steadystaterelation}). For the simple case we considered in Section \ref{sec:sstate}, in which magnetic fields are negligible (and, as is true throughout this paper, turbulence is also negligible), the steady-state assumption is valid for stars less than $\sim 0.4 M_\odot$ for fiducial molecular cloud parameters (Equation \ref{eqn:maxmass}). 

Sub-grid particle accretion methods have been employed to model protostellar accretion in large-scale numerical simulations of molecular clouds. Our results should be of particular utility for extending the sub-grid accretion models in such codes. Previous work has used unmagnetized accretion rates even though the sink particles were moving through a magnetized medium, thereby overestimating the true accretion rate onto the particle. We outline our implementation of Equations (\ref{eqn:mdotabhpar}) and (\ref{eqn:mdotabhper}) in \Orion sink particles in Appendix \ref{sec:sgrid} and demonstrate that this implementation succeeds in reproducing the correct accretion rate even when the accretion length scale $\sim r_{\rm ABH}$ is not well resolved (Figure \ref{fig:orion2subgriderror}). However, it should be noted that our accretion rates apply to gas that is not turbulent, and so they do not include the reduction associated with vorticity \citep{2005ApJ...618L..33K}.

Finally, our results have implications for the theory of star formation. At present, there are two main paradigms for the formation of massive stars: gravitational collapse, in which stars form via the gravitational collapse of a pre-existing protostellar core \citep{mckeetan03}, and competitive accretion, where protostars compete for gas from a common reservoir initially unbound to the stars \citep{zinnecker82,bonnell97,bonnell01}. Our study shows that magnetic fields make competitive accretion scenarios for the growth of pre-main sequence stars less efficient than predicted from Bondi-Hoyle accretion rates. For example, the amount of suppression for a cloud with average values of $\beta \sim 0.04$ \citep{crutcher99} and ${\cal M}\sim 1/2$ \citep{bonnell01} is a factor of $\sim20$ (Figures \ref{fig:mdotpar} and \ref{fig:mdotperp}). This reduction increases for lower $\b$ and ${\cal M}$. Models that employ Bondi accretion to transform molecular clouds into stars \citep[e.g.,][]{murraychang12} may be underestimating the timescale for the buildup of massive stars, and therefore, assuming these massive stars are what eventually destroy the cloud, are underestimating the lifetime of the molecular clouds in these models. Delayed buildup from magnetic fields would predict that these models have molecular clouds that persist beyond their typically observed lifetimes. In case of direct simulations, as long as the Alfv\'en-Bondi-Hoyle radius is resolved and the flow transitions to super-Alfv\'enic speeds, the accretion rates obtained will still be correct regardless of the subgrid model \citep[e.g,][also see the Appendix]{priceetal12}.

\acknowledgements
The authors thank the peer reviewer for an insightful report that helped improve the general clarity of the paper. The authors gratefully acknowledge support from (1) the National Science Foundation: A.T.L. through an NSF Graduate Fellowship, and C.F.M. and R.I.K. through grant AST-0908553 and AST-1211729, (2) the U.S. Department of Energy at the Lawrence Livermore National Laboratory: A.T.L. through grant LLNL-B569409, and R.I.K. and A.J.C. under contract DE-AC52-07NA27344, and (3) NASA: C.F.M. and R.I.K. through ATFP grant NNX13AB84G. Supercomputing support was provided through the NSF XSEDE at the University of Texas at Austin.

%=-=-=-=-=-=-=-=-=-=-=-=-=-=-=-=-

\appendix

\section{SUB-GRID MODEL}\label{sec:sgrid}

A collapsing molecular cloud forms structures that can be several to many orders of magnitude smaller in size than the original cloud. Large scale astrophysical simulations attempt to resolve these structures in order to follow their evolution, but this requires overcoming additional computational burdens, be it the reduced time step needed to ensure numerical stability or resource demands due to the increased memory requirements of the computational domain. Sink particle methods have been developed for both Lagrangian \citep{bate,hubberEtAl2013} and Eulerian mesh \citep{krumholzsink,federrath} codes to allow for collapsing flows that proceed beyond the finest resolved scale of the simulation. Material that enters these sink particles can be removed from the computational domain according to an analytical prescription that is intended to best estimate the physical processes that occur at those unresolved scales. In this section, we develop an implementation for embedding Lagrangian sink particles into an Eulerian mesh to model the accretion of an ideally magnetized gas, extending the approaches of \cite{krumholzsink,krumholzetal06} for non-magnetized flow. We review the criterion for the creation of a sink particle in a magnetized medium and we then determine the accretion rate of the sink particle, allowing for the finite resolution of the data. Only these prescriptions depend on the strength of the local magnetic field, whereas the others (sink particle mergers, coupling the sink particle's gravity to the hydrodynamics, etc.) do not and so are left unchanged. 
It should be borne in mind that, based on the observation that stars have far less magnetic flux than the gas from which they formed, we assume
that the sink particles accrete mass but not magnetic flux.

\subsection{Sink Particle Creation}

In simulations of gravitational collapse, mass accumulates in a small fraction of the grid cells. Sink particle algorithms must be able to identify whether these regions would continue to collapse if they were afforded higher resolution. On physical grounds, \citet{jeans02} showed that perturbations on scales larger than the Jeans length,
\beq \label{eqn:jeans} \lambda_{\rm J} = \left(\frac{\pi c^2_{\rm s}}{G\rho}\right)^{1/2}\ ,
\eeq
are unstable since thermal pressure cannot resist the self-gravity of the gas. \cite{truelove} showed that Eulerian simulations are subject to purely numerical fragmentation if this Jeans scale is not resolved by at least four cells. A sink particle is introduced at the center of a cell when the cell mass density $\rho$ exceeds 
a critical density, which we term the Truelove-Jeans density,
\beq \rho_{\rm TJ} = \frac{\pi J^2 c^2_{\rm s}}{G\,\Delta x^2} \ ,
\eeq
where $J = \Delta x / \lambda_{\rm J}$ is the (user-provided) inverse of the number of cells required resolve the local Jeans length. Once this is satisfied, a sink particle is initialized with mass $(\rho -\rho_{\rm TJ}) \Delta x^3$, and an equal mass is removed from the gas in the host cell. \citet{irdcpaper} extended this to incorporate ideal MHD, deriving a magnetic Truelove criterion: sink particles are initialized in cells whose density exceeds
\begin{equation}\label{eqn:magTL}
\rho_{\rm TJ,mag} = \rho_{\rm TJ}\left(1+0.74/\beta\right)\ 
\end{equation}
(see their Appendix A). The additional term arises in Equation (\ref{eqn:magTL}) because of the inclusion of magnetic pressure, which also acts to prevent gravitational collapse. \citet{federrath} arrived at a similar condition. Following the work of \citet{irdcpaper}, we adopt $J=1/8$ in our Truelove criterion for sink particle creation.

\subsection{Sink Particle Accretion}

\subsubsection{Estimated Accretion Rate, $\dot M_{\rm fit}$}

After the sink particle has formed, it will continue to accrete nearby gas. The rate of accretion onto the sink particle may be determined by processes that occur on scales that cannot be resolved in many modeling applications of interest. Here we develop an expression for the accretion rate onto sink particles that incorporates our new interpolation formulas (Equations \ref{eqn:mdotabhpar} and \ref{eqn:mdotabhper}) and that works at all resolutions. Our results for the accretion rate are based on the assumption that the ambient medium is uniform, but in simulations of star-forming regions, the medium is far from uniform. We therefore need an expression for the accretion rate that depends on locally measured quantities. To obtain this, we ran the models in this work and in \citet{cunnetal12} at a range of different resolutions and 
developed a prescription for the correct accretion rate based on quantities measured in the vicinity of the sink particle.

Two limits for the sink-particle accretion rate may be considered: The well-resolved case, in which $\rabh$ is much larger than the cell size, and the under-resolved case, in which it is much smaller. We need a prescription for the accretion rate that works in these limits, as well as in the intermediate case. Since $\rabh\propto M_*$, more massive stars are likely to have accretion flows that are well resolved, whereas low-mass stars are likely to have under-resolved flows
(see Section \ref{sec:sstate}).

First consider the well-resolved case. In this case, the
exact prescription for how much mass gets removed from the host and neighboring cells of the sink particle is not important since the flow's transition to velocities exceeding the fast magnetosonic velocity, $v_{\rm F}$, is resolved. 
Once $v \ge v_{\rm F}$, the flow becomes causally disconnected from the background and will collect near the sink particle regardless of the conditions of the surrounding medium. If our prescription underestimates the mass accretion rate, gas will collect in the sink particle's host cell until the gas density exceeds $\rho_{\rm TJ,mag}$, at which point a sink particle will be formed and immediately merged with the existing sink particle \citep{krumholzsink}. If our prescription overestimates the accretion rate, the density in the superfast infall will drop below the correct value, and the accretion rate in the next time step will be reduced. Thus, in the well-resolved case, the mass accretion rate is effectively set by the supersonic infall. The work described earlier in this paper using \Ramses is an example of this regime, since we ensured that the length scale $r_{\rm ABH}$ was resolved. The insensitivity to the accretion rate algorithm is also true for simulations of global supersonic collapse onto a particle \citep{shu}. 

Next, consider the under-resolved regime, where $r_{\rm ABH}$ is not well resolved. In this regime, the flow inside the sink cell is 
causally connected to the rest of the flow for $v_0<\vf$ (the subfast case) and to the flow in
the downstream Mach cone for $v_0>\vf$ (the superfast case).
The prescription for the amount of gas to be taken from the particle's host cell is important in this regime: not all the gas that flows through the cell should necessarily accrete onto the sink particle. Furthermore, the amount of gas in the particle's host cell determines the pressure support in the cell. The correct accretion rate in this regime is the Alfv\'en-Bondi-Hoyle rate that we have determined. The problem is that this accretion rate depends on the properties of the ambient medium, which we have assumed is homogeneous; in a simulation of a star-forming region, however, there is no homogeneous ambient medium. We therefore must estimate the accretion rate from the values of the parameters in the vicinity of the sink particle. Star-forming regions are supersonically turbulent, and
\citet{krumholzetal06} showed that such turbulence has two countervailing effects on the accretion rate: the rate is increased by the density fluctuations, but decreased by the vorticity in the flow. Our simulations include the first effect so long as density variations are well-resolved (i.e., except in shocks). We have not included vorticity, however, so our results are necessarily approximate when applied to a turbulent medium.

The results we have obtained for the accretion rate depend on quantities---the ambient density, $\rho$, 
the ambient plasma $\beta$, the Mach number of the flow past the accretor, $\calm$, and the angle
between the flow velocity and the field, $\theta$---that are
assumed to be constant far from the accreting particle. In star-forming regions, however, these 
quantities are not constant far from the accretor. To deal with this problem in simulations, we have developed a
two-step procedure: We first measure these quantities near the accretor; these values are denoted with a bar (e.g., $\bar\rho$).
In the limit of low-resolution, these values can be used for the ambient values that appear in our results for the accretion rate
in Section \ref{ssec:fitssection}. However, when the flow is resolved, gravitational focusing amplifies the density and magnetic field near the accretor, and values of these quantities must be extrapolated in order to estimate the ambient values; these estimates for the ambient values are denoted by a dagger (e.g., $\rho^\dag$). We do not distinguish the local and extrapolated values of the Mach number of the flow relative to the sink particle since they are the same in the limit of steady accretion (i.e., $\calm^\dag\simeq\bar\calm$).

In general, the angle $\theta$ between $\bvv$ and $\bB$ may be time-dependent, so it is not possible
to infer the ambient value of $\theta$ by extrapolation; we therefore set $\theta^\dag=\bar\theta$.
With Equations (\ref{eqn:mdotabhpar}) and (\ref{eqn:mdotabhper}) for $\dot{M}_{\|}$ and $\dot{M}_\perp$, our 
estimate of mass accretion rate onto a sink particle is then
\beq\label{eqn:sinkfit} 
\dot{M}_{\rm fit} = \dot{M}_{\perp}(\rho^\dag,\beta^\dag,\bar{\cal M}) \sin^2 \bar\theta + \dot{M}_{\|}(\rho^\dag,\beta^\dag,\bar{\cal M}) \cos^2 \bar\theta \ .
\eeq

In order to compute local values of the quantities $\bar\rho,\ \bar\beta,\ \bar{\cal M}$, and $\bar\theta$, we take averages over all the cells in a spherical shell of radius $\ravg\pm\Delta x$ around the sink particle,
where $\Delta x$ is the smallest grid size of the computational domain.  
The shell radius must be larger than $4\, \Delta x$, the region from which mass is removed from the cells and deposited onto the particle \citep{krumholzsink}, but not so large that the sink particle is sampling far from the local region that sets its current accretion rate. We have found the best results by sampling over a shell with radius $\ravg = 11\Delta x$. In particular, the value of $\bar{\cal M}$ is computed as a mass-average over the volume of the shell, and $\bar\theta$ is the angle between the volume-averaged magnetic field and volume-averaged momentum directions within this shell. We have found the best results by imposing the refinement criterion that every AMR level cover a sphere with a radius of 16 zones centered on the sink particle.

\begin{figure}
\plotone{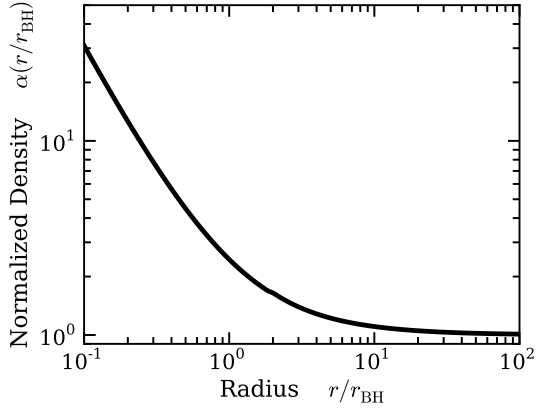}
\caption{Density profile from \citet{krumholzsink} for steady-state Bondi accretion onto a point source as a function of distance from the source $r$. The density is normalized to $\rho^\dag = \rho(r\rightarrow\infty)$.}
\label{fig:alpha}
\end{figure}

As noted above, if the flow near the sink is well resolved---i.e., if $\ravg< r_{\rm AB},\;r_{\rm BH}$---then the density and magnetic field will be amplified by gravitational focusing, so that $\bar\rho>\rho^\dag$ and $\bar B>B^\dag$. First consider the density. Following \citet{krumholzsink}, we determine $\rho^\dag$ by assuming that the density near the sink particle is well approximated by the stationary \citet{bondi52} solution, $\rho(r) = \rho^\dag\alpha(r/\rb)$, where the function $\alpha$ is determined by a set of transcendental equations. We evaluate this function at $r_{\rm avg}$, so that $\rho(r_{\rm avg})=\bar\rho$. To incorporate relative motion between the sink particle and the gas, we instead normalize the radius in $\alpha$ to $r_{\rm BH}$ \citep{krumholzsink}, giving
\beq
\bar\rho=\rho^\dag\alpha(r_{\rm avg}/r_{\rm BH})\ .
\eeq
The function $\alpha(r)$ is a monotonically decreasing function with the limit $\alpha\rightarrow 1$ as $r\rightarrow\infty$. In Figure \ref{fig:alpha} we plot $\alpha(r/\rbh)$. In the absence of a magnetic field, our estimate for the ambient density would be $\rho^\dag=\bar\rho/\alpha(\ravg/\rbh)$. Since the magnetic field limits the compression, we adopt an ansatz for the ambient density in which $\rho^\dag= \bar\rho \alpha^{-\chi}$, where $\chi$ goes smoothly from 1 in the hydrodynamic limit to 0 (i.e., no compression) in the limit of a strong field;
an explicit expression for $\chi$ as a function of $\beta^\dag$ will be given below. At sufficiently high resolution, directly simulating the accretion onto a totally absorbing sphere is more precise than our accretion rate fits.  We have found that this condition is met when 
$\Delta x \la \rabh/8$.
A resolution-dependent, piecewise prescription for the argument of $\alpha^{-\chi}$ is necessary to give a precise accretion rate in the limit of an asymptotically converged grid resolution, regardless of the error of our approximate fit.

Whatever the functional form for $\alpha^{-\chi}$, it has several requirements. In order to have the sink particle transition to a totally absorbing sphere in the high-resolution limit, the value of $\rho^\dag$ should be similar to the near-sink value of $\bar{\rho}$---
i.e., $\alpha^{-\chi}$ should be of order unity.
For under-resolved flows, $r_{\rm avg}$ is sampled farther from the sink particle. In the weak field limit, $\alpha^{-\chi}$ should approach $1/\alpha(r_{\rm avg}/r_{\rm BH})$, which gives an accurate correction factor between $\rho^\dag$ and $\bar\rho$ \citep{krumholzsink}. As $\beta$ decreases, the value of $\alpha^{-\chi}$ should decrease until the field becomes so strong that any gravitational enhancement of $\rho$ occurs well within $r_{\rm avg}$. 
For smaller values of $\beta$, $\alpha^{-\chi}$ should rise back up to be of order unity. 
We adopt
\beq\label{eqn:rhosink} 
	\rho^\dag = \bar{\rho} \left(\alpha\left[ \max\left(1,\frac{8\Delta x}{r_{\rm ABH}}\right) \frac{11 r_{\rm ABH}}{8 r_{\rm BH}}  \right]\right)^{-\chi}\ 
\eeq
as our functional form for $\rho^\dag$, which we later show achieves all the requirements above. Note that the factor 11/8 does not have any special significance; it is the result of our choice of
$\ravg=11\Delta x$ and our result that the criterion for being well resolved is $\Delta x<\rabh/8$.

The magnetic field is also amplified in the accretion flow.
For 1D compressions, $B\propto \rho$ so that $\beta\propto\rho/B^2\propto 1/\rho$. The accretion flow is far more complicated than that, but we use this simple relation as the basis for our ansatz for $\beta$. Guided by this asymptotic consideration, we choose the following ansatz for $\beta^\dag$:
\beq\label{eqn:betasink}
\beta^\dag=\bar{\beta} \left[\alpha \left(\dis\frac{11\Delta x}{r_{\rm BH}}\right)\right]^\chi .
\eeq
For the length scales $r_{\rm BH}$ and $r_{\rm ABH}$, the input quantity $\beta^\dag$ is used in the expressions given by Equations (\ref{eqn:rbh}) and (\ref{eqn:rabh}), so Equation (\ref{eqn:betasink}) an implicit function for $\beta^\dag$. Note that, in contrast to the prescription for $\rho^\dag$,  the prescription for $\beta^\dag$ does not require an explicit piecewise transition with resolution: the piecewise prescription for $\rho^\dag$ transitions our estimate to a totally absorbing sphere in the limit of high resolution, and by definition this is insensitive to the field strength.

It remains to give an expression for $\chi(\beta^\dag)$.
We construct this function so that the functional form of the density profile (Equation \ref{eqn:rhosink}) best reproduces the azimuthally-averaged steady state density profiles in \citet{cunnetal12}. The functions are fit to the profile in the equatorial plane, defined as the plane perpendicular to the original magnetic field direction that also goes through the center of the sink particle. After some experimentation, we obtained a reasonably good fit with 
\beq\label{eqn:chisink}
\chi(\beta^\dag) = \left\{
	\begin{array}{lcl}
	0 &:& \log_{10}\beta^\dag < {-3.1}, \\
	1.27-0.5/{(\beta^\dag)}^{\,0.13}  &:& {-3.1} \le \log_{10}\beta^\dag \le {2.0}, \\
	1 &:& \log_{10}\beta^\dag > {2.0}.
	\end{array}
\right.
\eeq
This function is a monotonically increasing function of $\beta^\dag$. With this final parameter specified, Equations (\ref{eqn:betasink}) and (\ref{eqn:chisink}) are solved simultaneously by iteration until $\beta^\dag$ converges to one part in $10^4$ or until $\beta^\dag > 10^9$. At this point, the four inputs $\rho^\dag,\ \beta^\dag,\ \bar{\cal M},$ and $\bar\theta$ are known and $\dot{M}_{\rm fit}$ can be determined. 

\begin{figure}
\plotone{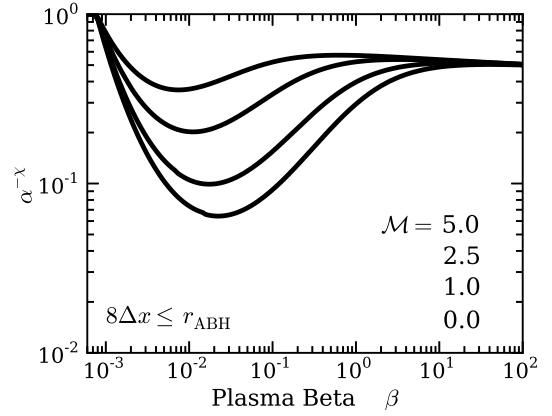}
\caption{The value of $\alpha^{-\chi}$ from Equation (\ref{eqn:rhosink}) for high resolution ($8\Delta x \le r_{\rm ABH}$). Four different Mach numbers are considered, increasing from bottom to top. The function for $\chi(\beta)$ is given in Equation (\ref{eqn:chisink}). For low resolution flows ($8\Delta x \ge r_{\rm ABH}$), $\alpha^{-\chi}$ is a monotonically decreasing function as $\Delta x$ decreases. Therefore, these curves also give the minimum value of $\alpha^{-\chi}$ for a particular $\beta$ and ${\cal M}$ in the low resolution limit. } 
\label{fig:alphchi}
\end{figure}

We now show that this formulation satisfies the criteria given above. First, in the high-resolution
limit, the sink should become totally absorbing, which requires that $\alpha^{-\chi}$ be of order
unity. In this case, the argument of $\alpha$ is $11\rabh/(8\rbh)$, so that 
$\alpha^{-\chi}\rightarrow[\alpha(11/8)]^{-1}\simeq 1/2$ in the high-$\beta$ limit. In the low-$\beta$ limit, $\chi\rightarrow 0$, so that $\alpha^{-\chi}\rightarrow 1$.
For intermediate values of $\beta$, $\alpha$ depends on both $\beta$ and $\calm$ as shown in
Figure \ref{fig:alphchi}. The smallest value of $\alpha^{-\chi}$ occurs for $\calm=0$; it is 0.064 at $\beta=0.022$. 

Next, for under-resolved flows ($\Delta x > r_{\rm ABH}/8$), $\alpha$ is evaluated at 
$\ravg/\rbh=11\Delta x/r_{\rm BH}$ and is resolution dependent. 
We argued that in the weak-field limit, $\alpha^{-\chi}$ should approach $[\alpha(\ravg/\rbh)]^{-1}$;
this occurs naturally, since $\chi\rightarrow1$ in this limit. In the strong-field limit,
we required $\alpha^{-\chi}\simeq 1$; this is satisfied since $\chi\rightarrow 0$ in this limit.
At intermediate values of $\beta$, we suggested that $\alpha^{-\chi}$ should have a minimum.
Although we have not portrayed $\alpha^{-\chi}$ for different values of $\Delta x$ in
Figure \ref{fig:alphchi}, this figure does show the expected minimum when $\Delta x$ is
at the boundary between low and high resolution ($\Delta x=\rabh/8$). As one moves into
the low-resolution regime (increasing $\Delta x$), the argument of $\alpha$ increases and
so does $\alpha^{-\chi}$. As a result, the values of $\alpha^{-\chi}$ for the high-resolution 
case in Figure \ref{fig:alphchi} provide a lower bound for the values in the low-resolution case.

\subsubsection{The Adjusted Accretion Rate, $\dot M_{\rm sink}$: Capping the Alfv\'en  Velocity}

With $\dot{M}_{\rm fit}$ given by the prescription above, mass is extracted from a sink region within $4\, \Delta x$ of the particle as an operator-split source term that is applied every fine AMR level time step increment $\Delta t$.  In well-resolved accretion flows care must be taken in extracting mass from the grid.  In such cases we do not want to introduce a new local maximum in the speed of magnetosonic waves---similar to what was described in Section \ref{sec:methods}.  In the opposite case of poorly-resolved accretion flow (e.g when $r_{\rm ABH}$ is not resolved) not introducing a new maximum in the value of $v_{\rm A}$, could arbitrarily diminish the accretion rate onto the sink particle. We therefore define a characteristic square-velocity $V^2$ as the maximum of two quantities, depending on whether $r_{\rm ABH}$ is resolved ($r_{\rm ABH} < \Delta x$) or not ($r_{\rm ABH} \ge \Delta x$):
\beq\label{eqn:sinkVsquared}
	V^2 = \max\left\{
		\begin{array}{lcl}
		\bar v^{\, 2}_{\, \rm A,max} &:& r_{\rm ABH} < \Delta x \\
		\dis\frac{\bar v^{\, 2}_{\, \rm A,max} \Delta x^2} {r_{\rm ABH}^2}\, &:&  r_{\rm ABH} \ge \Delta x
		\end{array}	
	\right.\ .
\eeq
The value for $\bar v^{\, 2}_{\, \rm A,max}$ is computed by taking the maximum Alfv\'en speed inside the same spherical shell described above. This defines the $\Delta\rho$ that can be extracted while only introducing a new local maximum in the Alfv\'en speed when $r_{\rm ABH}$ is not resolved and avoiding vanishing time-step pathologies when it is resolved. The value of $\Delta\rho$ extracted from a particular cell near the sink particle is set to the minimum of two quantities while holding the specific kinetic energy of the gas constant: 
\beq\label{eqn:deltarho}
	\Delta\rho = \min\left\{ 
	\begin{array}{lcl}
		(\dot{M}_{\rm fit}\Delta t/\Delta x^3)W(r) &:& \text{Mass accretion}\\ && \text{ estimated from} \\ \text{ fit.} \\
		\rho - B^2/(4\pi V^2) &:& \text{New maximum}\\ && \text{ Alfv\'en velocity} \\ && \text{avoided.}
	\end{array}      
	\right|\ ,
\eeq
where $r$ is the distance of that cell's center from the sink particle and the function $W(r)$ is a Gaussian kernel that extends out to $r=4\, \Delta x$ and is normalized to unity \citep{krumholzsink}. Note that the second expression for $\Delta\rho$ is 
non-negative
by definition. These two measures in the piecewise definition of $\Delta\rho$ are guided by physical considerations. However, if the sink particle accretes faster than the background flow can supply material, a void will open around the sink particle. 
If the density contrast between the void and the surroundings is allowed to become arbitrarily deep, the stability of the hydrodynamic scheme could be adversely impacted. This is particularly true in the limit of $\beta\rightarrow\infty$, where our Alfv\'en cap would not prevent this pathology. Therefore, we further impose the constraint that if the $\Delta\rho$ results in a particular cell having density less than $\rho^\dag/10$, then $\Delta\rho$ is adjusted so that the cell's density is floored at $\rho^\dag/10$. Finally, mass and momentum that is extracted from the grid is added to the sink particle mass in a manner than preserves global mass conservation,
\beq
	\Delta M_{\rm sink} = \sum_{ r\, \le\, 4\, \Delta x} \Delta\rho\, \Delta x^3\ ,
\eeq
and the momentum of the sink particle is updated in a likewise manner that preserves global momentum conservation.

\subsubsection{Verifying the Algorithm for the Accretion Rate}

\begin{figure}
\plotone{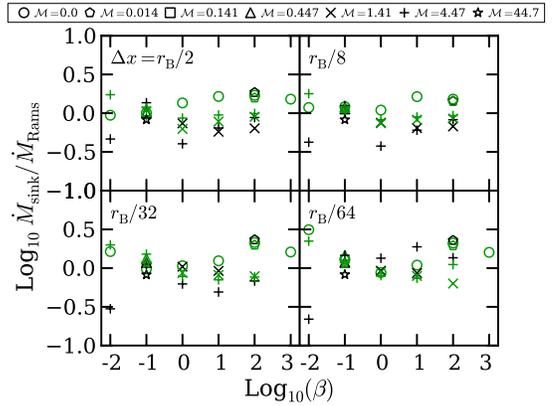}
\caption{Test of the magnetized sink particle algorithm implemented in \Orion. Plotted for each model is the ratio of steady-state accretion rates of \Orion $(\dot{M}_{\rm sink})$ and \Ramses $(\dot{M}_{\rm Rams})$ as a function of the smallest grid cell size $\Delta x$. Black points show the parallel orientation runs, green shows perpendicular. The Mach number of the run is given by the symbol.}
\label{fig:orion2subgriderror}
\end{figure}

\begin{figure}
\plotone{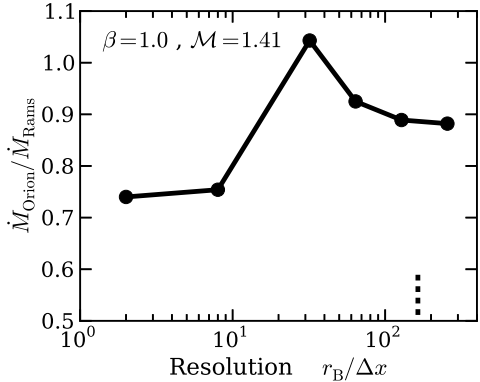}
\caption{Convergence study of the \Orion implementation as a function of the number of grid cells per $r_{\rm B}$. Plotted is the ratio of steady state accretion rates of \Orion versus \Ramses ($\dot{M}_{\rm Rams}$) for the ${\cal M}=1.41$, $\beta=1.0$ parallel case. For comparison, the default resolution \Ramses model had $r_{\rm B}/\Delta x \approx 164$, marked by the short dashed line. }
\label{fig:orion2convergence}
\end{figure}

We have implemented this MHD sink particle algorithm in the \Orion code \citep{orion2}.  To test the method, we repeat the models of this work and \cite{cunnetal12} on geometrically nested meshes having a base grid over $16 r_B$ of at least $64^3$ and enough AMR levels so that we coarsely resolve the accretion scales to be $r_{B}/\Delta x = 2,~8,~32,~{\rm and}\ 64$ on the finest level.  In Figure \ref{fig:orion2subgriderror} we show the comparison between the \Orion results and those from \Ramses. In general, we achieve accretion rates that are typically within a factor of two of the result obtained from high-resolution \Ramses\ models. In Figure \ref{fig:orion2convergence} we show the convergence properties in \Orion of a $\calm=1.41$, $\beta=1$ model with a parallel field orientation. The parameters of this test case were chosen so that magnetic, thermal and ram pressure effects are all of comparable importance and that the influence of all of these effects on $\dot{M}$ converge with sufficient resolution.  Note that at low resolution, the accretion rate is $\sim 40\%$ low and that the method transitions toward a pressure-less totally absorbing sphere by $r_{\rm B} / \Delta x > 64$, converging at high resolution to within $11\%$ of the \Ramses model. The $\sim11\%$ difference when \Orion is at comparable or higher resolution reflects intrinsic differences in the codes when resolving flows with a finite resolution. 

\bibliography{magbon}

%\begin{landscape}

% =-=-=-=-=-=-=-=-=-=-=-=-=-=-=-=-=-=-=-=-=-=-=-=-=-=-=-=-=-=-
% =-=-=-=-=-=-=-=-=-=-=-=-=-=-=-=-=-=-=-=-=-=-=-=-=-=-=-=-=-=-

\end{document}